\documentstyle[epsf,contractcite]{ioplppt}

   \eqnobysec
   \epsfclipon
   \newcommand{\smeq}{\! = \!}
   \newcommand{\smneq}{ \! \neq \!}

   \newcommand{\rmTr}{{\rm Tr}}
   \newcommand{\rmvar}{{\rm var}}
   \newcommand{\rmRe}{{\rm Re}}
   \newcommand{\rmIm}{{\rm Im}}
   \renewcommand{\fl}{\hspace*{-1.2cm}}
   \renewcommand{\lo}{\,}

\begin{document}

\title{Interference Phenomena in Electronic Transport Through Chaotic Cavities: 
       An Information-Theoretic Approach}

\author{Pier A. Mello\dag\ and Harold U. Baranger\ddag}

\address{
   \dag\ Instituto de F\'{\ii}sica, Universidad Nacional Aut\'onoma de M\'exico,
   01000 M\'exico D.F., M\'exico }

\address{
   \ddag\ Bell Laboratories-- Lucent Technologies,
    700 Mountain Ave. 1D-230, Murray Hill NJ 07974}

\bigskip
\submitted{1 December 1998}

\begin{abstract}
We develop a statistical theory describing quantum-mechanical scattering
of a particle by a cavity when the geometry is such that the classical
dynamics is chaotic. This picture is relevant to a variety of physical
systems, ranging from atomic nuclei to mesoscopic systems to microwave
cavities; the main application here is to electronic transport through
ballistic microstructures. The theory describes the regime in which there
are two distinct time scales, associated with a prompt and an equilibrated
response, and is cast in terms of the matrix of scattering amplitudes
$S$. The prompt response is related to the energy average of $S$ which,
through the notion of ergodicity, is expressed as the average over an
ensemble of similar systems. We use an information-theoretic approach: the
ensemble of $S$-matrices is determined by (1) general physical features--
symmetry, causality, and ergodicity, (2) the specific energy average of
$S$, and (3) the notion of minimum information in the ensemble. This
ensemble, known as Poisson's kernel, is meant to describe those situations
in which any other information is irrelevant. Thus, one constructs the
one-energy statistical distribution of $S$ using only information
expressible in terms of $S$ itself without ever invoking the underlying
Hamiltonian. This formulation has a remarkable predictive power: from the
distribution of $S$ we derive properties of the quantum conductance of
cavities, including its average, its fluctuations, and its full
distribution in certain cases, both in the absence and presence prompt
response. We obtain good agreement with the results of the numerical
solution of the Schr\"{o}dinger equation for cavities in which the
assumptions of the theory hold: ones in which either prompt response is
absent or there are two widely separated time scales. Good agreement with
experimental data is obtained once temperature smearing and dephasing
effects are taken into account.
\end{abstract}


\newpage

\tableofcontents

\newpage

\section{Introduction}
\label{introduction}

Scattering of waves by complex systems has captured the interest of
physicists for a long time. For instance, the problem of multiple scattering
of waves has been of great importance in optics \cite{newton,hulst}.
Interest in this problem has been revived recently, both for electromagnetic
waves \cite{Sheng} and for electrons \cite{AltLeeWebb91}, 
in relation with the phenomenon of localization,
which gives rise to a great many fascinating effects.

Nuclear physics, with typical dimensions of a few 
{\it f}m (1{\it f}m $\smeq 10^{-15}$ m), 
offers excellent examples of quantum-mechanical scattering by
``complex'' many-body systems dating as far back as the 1930's when
compound-nucleus resonances were discovered \cite{Feshbachbook}.
The treatment in these cases is often frankly statistical because the details 
of the many-body problem are intractably complicated. 

In the last two decades, electron transport in disordered metals has
been intensively 
investigated \cite{AltLeeWebb91,BvH91,LesHmeso,NATOmeso,Timpbook,BeenRMP}, 
as has transmission of electromagnetic waves through disordered 
media \cite{Sheng,AltLeeWebb91,LesHmeso}. 
The typical size scale is $1 \mu$m for electronic systems
and $1 \mu$m to $0.1$m in the electromagnetic case. 
Because these are also examples of scattering in complex environments, 
where the character of the disorder is not exactly known, a statistical 
approach which treats an ensemble of disordered potentials is natural.

Amazingly, features similar to those of these complex nuclear and disordered
problems have also been found in certain ``simple'' systems. 
While the geometry of these systems is apparently very simple-- 
quantum-mechanical scattering of just one particle by three circular disks 
in a plane, for instance-- the classical dynamics is fully chaotic. Such
systems have been studied by the ``quantum chaos'' community, in which the
main question is how the nature of the classical dynamics influences the
quantum properties \cite{LesHqchaos,Gutzbook}. In contrast to the complex
cases, these systems are amenable to exact (numerical) calculation; when the
results are analyzed statistically, however, the results are closely
related to those of the complex systems \cite{LesHqchaos,Gutzbook}. 
Experimentally, two types of simple 
scattering systems have been studied in particular: electron transport 
through microstructures called ``ballistic quantum dots'', whose dimensions
are of the order of 1$\mu $m, and microwave scattering from metallic
cavities, with typical dimensions of $0.1$m.
 
The ``universal'' statistical properties of wave-interference phenomena
observed in systems whose dimensions span about 14 orders of magnitude turn
out to depend on very general physical principles and constitute the central
topic of this article. Although our main interest throughout the paper is
electronic transport through ballistic chaotic cavities, or ballistic
quantum dots, described in section \ref{mesoscopics}, we wish to emphasize the
generality of the ideas involved, presenting first in
section \ref{nuclear physics} their application in the field of nuclear physics--
where some of them were first introduced-- with a brief reference to
microwave cavities at the end of that subsection.
 
\subsection{The atomic nucleus and microwave cavities}
\label{nuclear physics}
 
One of the most successful models in nuclear physics, called the optical
model of the nucleus, was invented by Feshbach, Porter and Weisskopf in the
1950's \cite{feshbach-et-al,feshbach}. That model, which works very
nicely over a wide range of energies, describes the scattering of a nucleon
by an atomic nucleus-- a complicated many-body problem-- in terms of two
distinct time scales:
 
1. A {\it prompt} response arising from {\it direct processes}, in which the
incident nucleon feels a mean field produced by the other nucleons. This
response is described mathematically in terms of the {\it average} of the
actual scattering amplitudes over an energy interval: these averaged
amplitudes, also known as {\it optical} amplitudes, show a much slower
energy variation than the original ones.
 
2. A {\it delayed}, or {\it equilibrated}, response, corresponding to the
formation and decay of the compound nucleus. It is described by the
difference between the exact and the optical scattering amplitudes: it
varies appreciably with energy and is studied with {\it statistical}
concepts using techniques known as random matrix theory
\cite{mehta,porter}.
 
Just as in the field of statistical mechanics time averages are very
difficult to construct and hence are replaced by {\it ensemble averages}
using the notion of {\it ergodicity}, in the present context too one finds
it advantageous to study energy averages in terms of ensemble averages
through an ergodic property
\cite{french-mello-pandey,agassi-et-al,lopez-mello-seligman,mello-seligman80}.
 
The optical model not only works well in nuclear physics, but has also been
applied successfully in the description of a number of chemical reactions,
thus bringing us from the nuclear to the molecular 
size scale \cite{levine}.
 
The connection of the above problems with the theory of waveguides and
cavities was proposed very clearly by Ericson and Mayer-Kuckuk more than 
thirty years ago \cite{ericson}:
``Nuclear-reaction theory is equivalent to the theory of waveguides... We
will concentrate on processes in which the incident wave goes through a
highly complicated motion in the nucleus... We will picture the nucleus as a
closed cavity, with reflecting but highly irregular walls.''
 
In fact, recent experiments with microwave cavities have shown features
similar to those that had been observed in the nuclear case. Importantly,
the ``irregular walls'' anticipated from the nuclear case are {\it not}
necessary in order to see these features: the analogy between nuclear
reaction theory and the theory of waveguides holds for simple smooth cavities
as long as the corresponding classical dynamics is chaotic
\cite{Gutz83,Blu88,SmilanRev}. This has been
the focus of several recent experiments involving microwave scattering
from metallic cavities \cite{Doron91,microwave1,microwave2,microwave3}.
These studies in simple chaotic systems were predated by extensive studies 
on scattering of microwaves by a disordered dielectric medium;
particularly important are the experiments of A. Genack and
collaborators \cite{genack}. It is also interesting to note that the use of
statistical concepts to analyze electromagnetic scattering by waveguides
started quite independently of the connection to the complex scattering
of nuclear physics, in the context of radio wave propagation \cite{radio}.
 
\subsection{Ballistic Mesoscopic Cavities}
\label{mesoscopics}
 
The term {\it mesoscopic } system refers to microstructures in which the
phase of the single-electron wave function-- in an independent-electron
approximation--  remains coherent across the system of interest
\cite{AltLeeWebb91,BvH91,LesHmeso,NATOmeso,Timpbook,BeenRMP}: this means that 
the phase-coherence length $l_{\phi }$ associated with processes that can change
the environment-- the other electrons, or the phonon field-- to an orthogonal
state exceeds the system size \cite{BvH91}. This is realized in the
laboratory with systems whose spatial dimensions are on the order of 1$\mu $m
or less and at temperatures $\leq$1 K. In so-called {\it ballistic cavities},
or {\it quantum dots}, the electron motion is in addition practically
ballistic, except for specular reflection from the walls: thus the elastic
mean free path $l_{\rm el}$ also exceeds the system size.
In the most favorable material system, GaAs heterostructures, this condition 
can also be realized for cavities of size $\leq 1 \mu$m
\cite{BvH91,LesHmeso,NATOmeso,Timpbook}.

Experimentally, an electrical current is established through the leads that
connect the cavity to the outside and the potential difference across the
cavity is measured, from which the conductance
$G$ is extracted. In an independent-electron picture one thus aims at
understanding the quantum-mechanical single-electron scattering by the
cavity, while the leads play the role of waveguides. It is the multiple
scattering of the waves reflected by the various portions of the cavity that
gives rise to interference effects.  Three important experimental probes of
the interference effects are an external magnetic field $B$, the Fermi energy
$\epsilon_F$, and the shape of the cavity: when these are varied the relative
phase of the various partial waves changes and so the interference pattern
changes. The changing interference pattern in turn causes the conductance to
change; this sensitivity of $G$ to small changes in parameters through
quantum interference is called {\it conductance fluctuations}.

The connection between scattering by simple chaotic cavities
\cite{Gutz83,Blu88,SmilanRev} and mesoscopic systems was first made
theoretically \cite{JalBarSto90}. Subsequently,
cavities in the shape of a stadium, for which the single-electron classical
dynamics would be chaotic, were first reported in Ref. \cite{Marcus93}. 
More recently, several other types of structures have been investigated
including experimental {\it ensembles} of shapes
\cite{Chang,Westervelt,Bird,Prober,Marcus95,Weis,Ensslin,Mailly,Sachrajda,%
Swedish,Marcus98}. Averages of the conductance, its
fluctuations, and its full distribution were obtained over such an ensemble.

It is the aim of the present paper to provide a theoretical framework to
describe this physical situation. To this end we will set up a scheme
similar to that explained in section \ref{nuclear physics} in connection with
the scattering problem in nuclear physics. A complementary treatment based
on semiclassical ideas has also been developed but will not be covered
here \cite{JalBarSto90,SemiclRev}.

The paper is organized as follows. We start by presenting the general ideas
of quantum-mechanical scattering by a cavity, introducing the scattering or
$S$ matrix of the problem (Sec. \ref{scatt}), and then turn to 
treating an
ensemble of systems in terms of an ensemble of $S$ matrices 
(Sec.~\ref{ensembles}). 
Specific analytical results for the conductance are then
presented, first in the absence of direct processes (Sec. \ref{no direct
processes}), and then with direct processes present (Sec. \ref{direct
processes}). We then compare our theoretical results with the numerical
solution of the Schr\"{o}dinger equation (Sec. \ref{numerics}), and finally
in Sec. \ref{dephasing} compare with the experimental data that were already
mentioned above. In the latter comparisons, a number of discrepancies are
found; to reconcile theory and experiment, we realize the necessity of
introducing the effect of processes that destroy the coherence of the
wave function in the sample. Finally, the conclusions of our work are
presented in Sec. \ref{conclusion}. Some of the main results in this paper
have appeared in condensed form in our previous publications
\cite{BarMelloPRL94,BarMello95,BarMelloEPL96}.

\section{The scattering problem}
\label{scatt}

\subsection{Scattering waves: Definition of the $S$-matrix}
\label{scattwaves}

\begin{figure}[tb]
   \begin{center}
   \leavevmode
   \epsfxsize=8.0cm
   \epsffile[50 285 590 770]{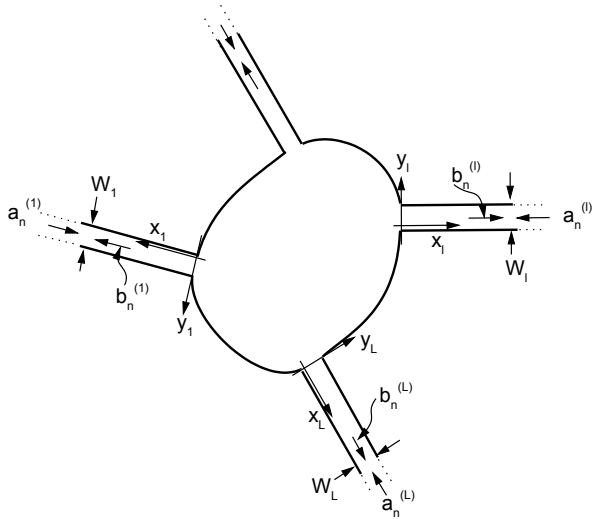}
   \end{center}
   \vspace*{-0.5cm}\caption{The 2D cavity studied in the text. 
   The cavity is connected to the 
   outside via $L$ waveguides. The arrows inside the waveguides indicate 
   incoming or outgoing waves as in Eq. (\protect\ref{build. block sols.}). 
   In waveguide $l$ there can be $N_l$ such incoming or outgoing waves: this 
   is indicated in the figure by the amplitudes $a^{l}_n$, $b^{l}_n$, 
   respectively, where $n \smeq 1,\cdot \cdot \cdot ,N_l$.}
   \label{cavity}
\end{figure}

Consider a system of noninteracting electrons. Since we shall be dealing
with cases in which spin-orbit coupling is negligible, we disregard the spin
degree of freedom and just consider ``spinless electrons'' in what follows.
We are interested in studying the scattering of an electron at the Fermi
energy $\epsilon _F \smeq \hbar ^2k_F^2/2m$ by the 2D microstructure shown
schematically in Fig. \ref{cavity}. 
The microstructure consists of a cavity,
connected to the outside by $L$ leads, ideally of infinite length, that play
the role of waveguides. The $l$-th lead ($l \smeq 1,\cdot \cdot \cdot ,L$) has
width $W_l$. We are interested in the (scattering) solutions of the
Schr\"odinger equation {\it inside} such a structure, with the ideal
boundary condition that the walls of the cavity and leads are completely
impenetrable: hence the wave function must vanish there.

In each lead $l$ we introduce a system of coordinates $x_l$, $y_l$, as
indicated in Fig. \ref{cavity}. The $x_l$ axis runs along the lead and
points outwards from the cavity. The $y_l$ axis runs in the transverse
direction and is tangential to the cavity wall and its continuation across
the lead; $y_l$ takes on the values $0$ and $W_l$ on the two walls of the
lead. In lead $l$ and for $x_l>0$ we have the elementary solutions to the
Schr\"odinger equation 
\begin{equation}
e^{\pm ik_n^{(l)}x_l}\chi _n(y_l)  \label{build. block sols.}
\end{equation}
where the positive (negative) sign is for outgoing (incoming) waves.
Here the functions $\chi _n(y_l)$ are the solution of the transverse
Hamiltonian which in the presence of a magnetic field may depend on
$k_n^{(l)}$ \cite{BarSto89}. 
The solution of the scattering problem consists in expressing
the amplitude of the outgoing waves in terms of the incoming ones.

In the absence of a magnetic field, the explicit problem can be simply stated; 
we now present this case in detail, noting that it can be generalized to the 
$B \smneq 0$ case \cite{BarSto89}. For $B \smeq 0$, 
the functions $\chi _n(y_l)$ are 
\begin{equation}
\chi _n(y_l)=\sqrt{\frac 2{W_l}}\sin K_n^{(l)}y_l, \quad 
K_{n}^{(l)}=\frac{n\pi }{W_{l}},\quad n \smeq 1,2,\ldots 
\label{chi n}
\end{equation}
where $K_n^{(l)}$ is the ``transverse wave number''.
The functions $\chi _{n}(y_{l})$ vanish on the two walls of the lead and
form a complete orthonormal set of functions for the variable $y_{l}$; i.e. 
\begin{equation}
\left\langle \chi _{n}|\chi _{n}\right\rangle =\delta _{nm}.
\label{orthonormality}
\end{equation}
This ``transverse quantization'' is a consequence of the boundary condition
on the walls of the leads. Each possibility defined by the integer $n$ is
named a {\it mode}, or {\it channel}. The ``longitudinal'' wave number $%
k_{n}^{(l)}$ satisfies the relation 
\begin{equation}
[ k_{n}^{(l)}] ^{2}+[ K_{n}^{(l)}] ^{2}=k_{F}^{2}.
\label{def of longit  k}
\end{equation}
If $K_{n}^{(l)}<k_{F}$, then $[ k_{n}^{(l)}] ^{2}>0$, 
$k_{n}^{(l)}$ is real, and the $e^{\pm ik_{n}^{(l)}x_{l}}$ occurring in 
Eq. (\ref{build. block sols.}) represent running waves along the leads: we thus
have {\it running modes} or {\it open channels}. On the other hand, when 
$K_{n}^{(l)}>k_{F}$, then $[ k_{n}^{(l)}] ^{2}<0$ and $k_{n}^{(l)}$ is pure 
imaginary, thus giving rise to exponentially decaying
waves along the leads: these modes are called {\it evanescent modes} or 
{\it closed channels}. If 
\begin{equation}
N_{l}<k_{F}W_{l}/\pi <N_{l}+1,  \label{Nl open channels}
\end{equation}
there are $N_{l}$ open channels in lead $l$. Very far away along the leads,
i.e. as $x_{l}\rightarrow \infty $, only the contribution of the open
channels contributes to the wave function. The most general form of the
asymptotic wave function in lead $l$ is thus the linear combination 
\begin{equation}
\sum_{n=1}^{N_{l}}\left[ a_{n}^{(l)}\frac{{\rm e}^{-ik_{n}^{(l)}x_{l}}}{%
(\hbar k_{n}^{(l)}/m)^{1/2}}+b_{n}^{(l)}\frac{{\rm e}^{ik_{n}^{(l)}x_{l}}}{%
(\hbar k_{n}^{(l)}/m)^{1/2}}\right] \chi _{n}(y_{l}).
\label{gen wf in lead l}
\end{equation}
Note that the normalization of the plane waves is such that they give rise to 
{\it unit flux}.

We define the $N_l$-dimensional vector 
\begin{equation}
{\bf a}^{(l)}=
( a_1^{(l)}, \ldots, a_{N_l}^{(l)} )^T
\label{al}
\end{equation}
that contains the $N_l$ incoming amplitudes in lead $l$ 
($l \smeq 1,\ldots,N_l$).
Putting all the ${\bf a}^{(l)}$ ($l \smeq 1,\ldots,L$) together, we form the 
vector 
\begin{equation}
{\bf a}= ( {\bf a}^{(1)}, \ldots, {\bf a}^{(L)} )^T
\label{a}
\end{equation}
associated with the incoming waves in all the channels for all the leads. We
can make similar definitions for the outgoing-wave amplitudes. The {\it %
scattering matrix}, or $S${\it \ matrix}, is then defined by the relation 
\begin{equation}
{\bf b}=S{\bf a,}  \label{bSa}
\end{equation}
connecting the incoming to the outgoing amplitudes. In terms of individual
leads we can write 
\begin{equation}
S=\left[ 
\begin{array}{cccc}
r_{11} & t_{12} & \ldots & t_{1L} \\ 
t_{21} & r_{22} & \ldots & t_{2L} \\ 
\vdots & \vdots & \ddots & \vdots \\ 
t_{L1} & t_{L2} & \ldots & r_{LL}
\end{array}
\right] .  \label{S}
\end{equation}
Here, $r_{ll}$ is an $N_l\times N_l$ matrix, containing the reflection
amplitudes from the $N_l$ channels of lead $l$ back to the same lead; $%
t_{lm} $ is an $N_l\times N_m$ matrix, containing the transmission
amplitudes from the $N_m$ channels of lead $m$ to the $N_l$ channels of lead 
$l$. The $S$ matrix is thus a square matrix with dimensionality $n$ given by 
\begin{equation}
n=\sum_{l=1}^LN_l.  \label{n}
\end{equation}
While we have carried out the analysis explicitly for $B \smeq 0$, the
main results, Eqs. (\ref{build. block sols.}), (\ref{gen wf in lead l}),
(\ref{bSa}), and (\ref{S}), also hold in the presence of a magnetic field.

Having chosen the unit-flux normalization for the plane waves in Eq. (\ref
{gen wf in lead l}), {\it flux conservation} implies {\it unitarity} of the $%
S$ matrix \cite{newton,kirk,lane,merzbacher}; i.e. 
\begin{equation}
SS^{\dagger }=I.  \label{unitarity}
\end{equation}
In the absence of other symmetries, we have this requirement only. This is
the {\it unitary} case, also designated in the literature as $\beta  \smeq 2.$ 
In the presence of {\it time-reversal invariance} (TRI) (as in the
absence of a magnetic field) and no spin, the $S$ matrix, besides being
unitary, is{\it \ symmetric} \cite{newton,kirk,lane,merzbacher,dyson62}: 
\begin{equation}
S=S^{{T}}  \label{symmetry}
\end{equation}
This is the {\it orthogonal} case, also designated as $\beta  \smeq 1$. The 
{\it symplectic} case ($\beta  \smeq 4$), arising in the presence of spin and
time-reversal invariance, will not be touched upon in this presentation.

\subsection{ The conductance}
\label{conductance}

For a two-lead problem, which will occur most frequently in our analysis,
the $S$ matrix has the structure 
\begin{equation}
S=\left[ 
\begin{array}{cc}
r_{11} & t_{12} \\ 
t_{21} & r_{22}
\end{array}
\right] \equiv \left[ 
\begin{array}{cc}
r & t^{\prime } \\ 
t & r^{\prime }
\end{array}
\right] .  \label{S n=2}
\end{equation}
In the particular case $N_1 \smeq N_2 \smeq N$, i.e. when the two leads have 
the same number of channels $N$, the four blocks $r$, $t$, $r^{\prime }$, 
$t^{\prime } $ are $N\times N$ and the $S$ matrix is $2N\times 2N$.

As we mentioned in the Introduction, we are interested in the electronic 
{\it conductance} of the microstructure. If we assume that the latter is
placed between two reservoirs (at different chemical potentials) shaped as
expanding horns with negligible reflection back to the microstructure, then
the Landauer-B\"uttiker formula \cite{buettiker,levinson,levinson-shapiro}
expresses the conductance $G$ in terms of the {\it scattering properties of
the microstructure itself} as 
\begin{equation}
G=\frac{e^2}hg \; , \;\;\; g=2T \; ,  
\label{G}
\end{equation}
where the factor $2$ arises from the two spin degrees of freedom and
the ``spinless conductance'' $T$ is the transmission coefficient 
$\left|t_{ab}\right| ^2$ summed over initial and final channels:
\begin{equation}
T={\rmTr}(tt^{\dagger }).  \label{T}
\end{equation}

\subsection{Polar representation of the $S$ matrix}
\label{parametrization S}

In the two-lead case with $N_1 \smeq N_2 \smeq N$, one can parametrize 
the $S$ matrix in the so called ``polar representation'' 
as \cite{hua,mello-pichard,MelPeyKumar,BarMelloPRL94} 
\begin{equation}
S=\left[ 
\begin{array}{cc}
v_1 & 0 \\ 
0 & v_2
\end{array}
\right] \left[ 
\begin{array}{cc}
-\sqrt{1-\tau } & \sqrt{\tau } \\ 
\sqrt{\tau } & \sqrt{1-\tau }
\end{array}
\right] \left[ 
\begin{array}{cc}
v_3 & 0 \\ 
0 & v_4
\end{array}
\right] 
=VRW \;\; .  
\label{S polar repr}
\end{equation}
Here, $\tau $ stands for the $N$-dimensional diagonal matrix of eigenvalues 
$\tau _a$ ($a \smeq 1,\cdot \cdot \cdot ,N$) of the Hermitian matrix 
$tt^{\dagger }$. The $v_i$ ($i \smeq 1,\cdot \cdot \cdot ,4$) are arbitrary 
$N\times N$ unitary matrices for $\beta  \smeq 2$, with the restrictions 
\begin{equation}
v_3=v_1^T,\qquad v_4=v_2^T  \label{cond for symm}
\end{equation}
for $\beta  \smeq 1$. It is readily verified that any matrix of the form (\ref{S
polar repr}) satisfies the appropriate requirements of an $S$ matrix for $%
\beta  \smeq 1,2$. The converse statement, as well as the uniqueness of the
polar representation, can also be proved using an argument similar to
that in Ref. \cite{mello-pichard}.

In the polar representation, we can write the total transmission $T$  of
Eq. (\ref{T}) as
\begin{equation}
T=\sum_{a}\tau _{a}.  \label{g(tau)}
\end{equation}
Thus the polar representation is natural for the study of conductance
since it separates the magnitude of the transmission-- the transmission
eigenvalues $\{ \tau_a \}$-- from the irrelevant phase effects-- the unitary
matrices $\{v_i \}$.

\section{Ensembles of $S$ matrices: An information-theoretic approach}
\label{ensembles}

As we mentioned in the Introduction, we are interested in ensembles of
systems, which will be represented as {\it ensembles of }$S${\it \ matrices}
endowed with a {\it probability measure}. For that purpose it is important
to introduce first the notion of {\it invariant measure }for our $S$
matrices: this we develop in the first subsection. The second subsection is
devoted to the derivation of the probability measure for our ensemble of $S$
matrices, using an information-theoretic point of view.

\subsection{The invariant measure}
\label{invariant measure}

The invariant measure is the measure which equally weights all matrices 
which satisfy the unitarity and symmetry constraints. Intuitively, it 
corresponds to the most random distribution consistent with the constraints, 
or in other words the one with the least information.
Mathematically, such a measure is defined by requiring that it 
remain invariant under an automorphism of a
given symmetry class of matrices into itself \cite{dyson62,hua}:
\begin{equation}
d\mu ^{(\beta )}(S)=d\mu ^{(\beta )}(S^{\prime }).  \label{inv measure}
\end{equation}
For $\beta  \smeq 1$, the transformed matrix $S^{\prime }$ is related to $S$ by 
\begin{equation}
S^{\prime }=U_{0}SU_{0}^{T},  \label{autom orthogonal}
\end{equation}
$U_{0}$ being an arbitrary, but fixed, unitary matrix. Clearly, 
Eq. (\ref{autom orthogonal}) is an automorphism of the set of unitary symmetric
matrices into itself. For $\beta  \smeq 2,$ 
\begin{equation}
S^{\prime }=U_{0}SV_{0}  \label{autom unitary}
\end{equation}
where $U_{0{\ }}$ and $V_{0}$ are arbitrary fixed unitary matrices. For 
$\beta  \smeq 2$, the resulting measure is the well known Haar's measure of the
unitary group and its uniqueness is well known \cite{wignerbook,hamermesh}.
Uniqueness for $\beta  \smeq 1$ was shown in Ref. \cite{dyson62}. 
Using the invariant measures, 
Eqs. (\ref{inv measure}) -  (\ref{autom unitary}),
as the probability measures for ensembles of $S$-matrices defines the
{\it Circular Orthogonal} and {\it Unitary Ensembles (COE, CUE)} for 
$\beta  \smeq 1,2$, respectively.

Several explicit representations of the invariant measure are known, the 
classic one being in terms of the eigenphases and eigenvectors of the
$S$-matrix. For our purposes, the polar representation of 
Eq. (\ref{S polar repr}) 
is of particular interest because of its connection to the conductance
properties of the cavity. We thus consider the two-equal-lead case, 
$N_1 \smeq N_2 \smeq N$, and express the invariant measure explicitly in this
parametrization.

We first recall a well known result from differential geometry. Consider the
expression for the arc element
\begin{equation}
ds^{2}=\sum g_{\mu \nu }(x)\delta x_{\mu }\delta x_{\nu },
\label{arc element}
\end{equation}
written in terms of {\it independent variables} and the metric tensor $%
g_{\mu \nu }(x)$. Assuming that $ds^{2}$ remains invariant under the
transformation $x_{\mu } \smeq x_{\mu }(x_{1}^{\prime },x_{2}^{\prime },\cdot
\cdot \cdot )$, one can prove that the volume element
\begin{equation}
dV=|{\rm det}g(x)|^{1/2}\prod_{\mu }dx_{\mu }  \label{vol element}
\end{equation}
remains invariant under the same transformation \cite{lass}.

We now go back to our random-$S$-matrix problem. We define the differential
arc element as 
\begin{equation}
ds^{2}= \rmTr  [dS^{\dagger }dS].  \label{arc element for S}
\end{equation}
This expression remains invariant under the transformations 
(\ref{autom orthogonal}) and (\ref{autom unitary}). 
Substituting for $S$ the form (\ref{S polar repr}), one can extract the 
metric tensor; applying Eq. (\ref{vol element}), one then finds the 
invariant measure.  This is done in 
\ref{app inv measure}; the result is \cite{BarMelloPRL94,carloEPL94}
\begin{equation}
d\mu ^{({\beta })}(S)=P^{({\beta })}(\{\tau \})\prod_{a}d\tau
_{a}\prod_{i}d\mu (v^{(i)}) \; ,  \label{inv meas (tau v)}
\end{equation}
where $P^{({\beta })}(\{\tau \})$ denotes the joint probability density of
the $\{ \tau \}$, 
$d\mu (v^{(i)})$ is the invariant or Haar's measure on the unitary group 
$U(N)$ \cite{hamermesh}, and $C_{\beta }$ is a normalization constant.
From \ref{app inv measure} we have
\begin{equation}
P^{(1)}(\{\tau \})=C_{1 }\prod_{a<b}\mid \tau _{a}-\tau _{b}\mid
\prod_{c}\frac{1}{\sqrt{\tau _{c}}} , \qquad
P^{(2)}(\{\tau \})=C_{2 }\prod_{a<b}\mid \tau _{a}-\tau _{b}\mid ^{2}.
\label{w(tau)}
\end{equation}
Eqs. (\ref{inv meas (tau v)}) and (\ref{w(tau)}) explicitly specify the
invariant measure. The factor involving the product over pairs
gives the repulsion of the eigenvalues; notice that the repulsion is linear
for the orthogonal case while quadratic in the unitary case,
as typically occurs.

\subsection{The information-theoretic model}
\label{information theoretic model}

\subsubsection{The one-channel case}

To begin our discussion, consider first a physical problem that can be
described by a $1\times 1$ $S$ matrix. This is the case, for instance, for a
particle scattered by a 1D potential that is nonzero in the region $-a\leq
x\leq 0$, to which an impenetrable wall is added at $x \smeq -a$: the
particle then lives in the semiinfinite domain $-a\leq x\leq \infty $.
Another example is that of a 2D cavity, connected to the outside by only one
lead that in turn supports only one open channel. From unitarity, $S$ must
be a complex number of unit modulus at every energy; i.e. $S(E) \smeq
e^{i\theta (E)}$. In the Argand diagram $\rmRe (S)$-$\rmIm (S)$, $S(E)$ is
represented, for a given energy, by a point on the unitarity circle: that
point is defined by the angle $\theta (E)$. As the energy changes, so does
the representative point: this is what we may call, pictorially, the
``motion'' of $S(E)$ as a function of energy; it resembles the motion in
phase space of the representative point of a classical system as a function
of time.

We ask the following question: as we move along the energy axes, {\it what
fraction of the time do we find }$\theta ${\it \ lying inside a given
interval }$d\theta ${\it ?} Let us call $dP(\theta ) \smeq p(\theta )d\theta$ 
that fraction.

To answer this question, we first analyze how to construct energy averages.
By this we always mean a {\it local} energy average, performed inside an
interval $I$ that contains many resonances and yet is small compared to an
energy interval over which ``gross structure'' quantities showing a secular
variation, like the average spacing $\Delta $ of resonances, vary
appreciably. We then expect the dependence of the average in question on the 
center of the interval $E_{0}$, its width $I$, as well as the particular 
weighting function used to define the average, to be {\it weak}.

We devise an {\it idealized} situation: the argument $E$ in $S(E)$ is
extended all the way from $-\infty $ to $+\infty $, in such a way that {\it %
local averages are everywhere the same as inside }$I$: this idealization,
which we call {\it stationarity}, will only represent well what goes on {\it %
locally} inside $I$ in the actual system.
We indicate an energy average of a quantity by placing a bar over it while an
ensemble average is denoted by angular brackets.

Ref. \cite{lopez-mello-seligman} chooses, as the weighting function, a
Lorentzian. Using the fact that $S(E)$ is {\it analytic} in the upper half
of the complex energy plane ({\it causality}), it shows that
\begin{equation}
\overline{S^{k}}=\left[ \, \overline{S}\, \right] ^{k},  \label{<Sk>=<S>k}
\end{equation}
i.e. {\it the average of the }$k${\it -th power of }$S$ {\it coincides with
the }$k${\it -th power of the average of }$S$. Thus, the quantity 
$\overline{S}$,
referred to as the optical $S$ matrix in Sec. \ref{nuclear physics}, 
plays a special
role, in that the average of any power of $S$ can be expressed in terms of
it. Ref. \cite{lopez-mello-seligman} then finds the answer to the
above-posed question: the fraction of time $p(\theta )$ spent by $\theta $
in a unit interval around $\theta $ in its journey along the energy axis is 
{\it uniquely} given by the expression 
\begin{equation}
p(\theta )=\frac{1}{2\pi }\frac{1-|\, \overline{S}\, |^{2}}
{|S-\overline{S}\, |^{2}},\qquad S=e^{i\theta }\;,  \label{poisson}
\end{equation}
and {\it depends only upon the average, or optical, }$S${\it \ matrix }
$\overline{S}$. 
This expression is also known as {\it Poisson's kernel} \cite{hua}. The
conclusion is remarkable: it tells us that the {\it system specific details
are irrelevant}, except for the optical $S$ matrix.

\begin{figure}[tb]
   \begin{center}
   \leavevmode
   \epsfxsize=9.0cm
   \epsffile[150 442 496 645]{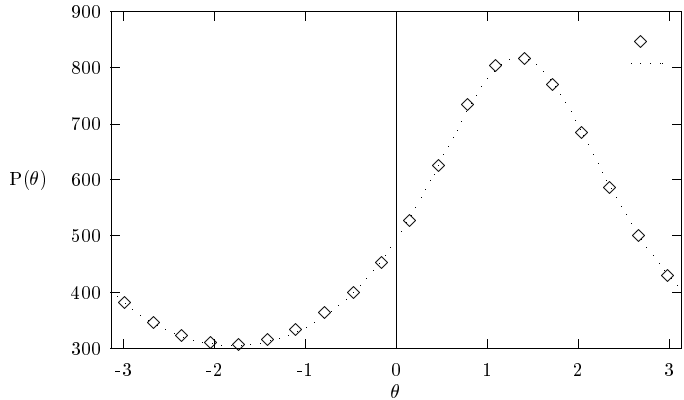}
   \end{center}
   \vspace*{-0.5cm}\caption{ In the $\delta $-potential model a stretch of 
   energy containing 100 resonances, starting from $ka \smeq 10,000$,
   was sampled to find the fraction of time that $\theta $ falls in a unit
   interval around $\theta $: the result is indicated by diamonds. The
   curve is a plot of Poisson's kernel (\protect\ref{poisson}), with the value
   of $\overline{S}$ extracted from the numerical data, in order to have a
   parameter-free fit. The agreement is excellent. 
   (From Ref. \protect\cite{victor}.)}
   \label{ptheta}
\end{figure}

As an example, consider a 1D $\delta $-potential centered at $x \smeq 0$ and a
perfectly reflecting wall at $x \smeq -a$. An energy stretch containing 100
resonances starting from $ka \smeq 10,000$ (so that secular variations of
gross-structure quantities can be neglected) was sampled to find the
fraction of time that $\theta $ falls in a certain small interval around $%
\theta $ \cite{victor}. The result is compared in Fig. \ref{ptheta} with
Poisson's kernel (\ref{poisson}), where the value of the optical $\overline{S}$
was extracted from the numerical data themselves, so as to have a
parameter-free fit. We observe that the agreement is excellent.

Consider now a collection of systems, described by an ensemble of $S$
matrices endowed with a probability measure. As an idealization, suppose we
further construct $S(E)$ as a {\it stationary random function of energy},
for $-\infty <E<\infty $. Then we know the conditions under which {\it %
ergodicity}, understood as equality of energy and ensemble averages (except
for a set of zero measure), holds \cite{yaglom}. Let us then assume that our
ensemble is ergodic.

The condition (\ref{<Sk>=<S>k}) arising from analyticity, together with
ergodicity, implies the relation
\begin{equation}
\left\langle S^{k}\right\rangle =\left\langle S\right\rangle ^{k}
\label{AE 1 ch.}
\end{equation}
between ensemble averages, often called the {\it analyticity-ergodicity (AE)}
requirement. The ensemble measure is thus {\it uniquely} given by
\begin{equation}
dP_{\left\langle S\right\rangle }(S)=p_{_{\left\langle S\right\rangle
}}(\theta )d\theta ,  \qquad
p_{_{\left\langle S\right\rangle }}(\theta )=\frac{1}{2\pi }
\frac{1-|\left\langle S\right\rangle |^{2}}
{|S-\left\langle S\right\rangle |^{2}} \; ,
\label{poisson ensemble n=1}
\end{equation}
once $\langle S \rangle $ is specified. The ensemble depends parametrically 
upon the single complex number $\langle S \rangle $, {\it any other 
information being irrelevant}.

We note in passing that Eq. (\ref{AE 1 ch.}) implies that a function $f(S)$
that is analytic in its argument, and can thus be expanded in a power series
in $S$, must fulfill the {\it reproducing property } \cite
{hua,mello-pereyra-seligman}
\begin{equation}
f(\left\langle S\right\rangle )=\int f(S)dP_{\left\langle S\right\rangle
}(S).  \label{repr. prop.}
\end{equation}
It is because the probability measure appears as the kernel of this integral
equation that it is called Poisson's kernel.

\subsubsection{The multi-channel case}

We now consider $S$ matrices of dimensionality $n$, that can describe, in
general, a multi-lead problem with $n$ channels altogether, as explained in
Sec. \ref{scatt}. The Argand diagram discussed above for $n \smeq 1$ has to be
generalized to include the axes $\rmRe S_{11}$, $\rmIm S_{11}$, $\rmRe S_{12}$, 
$\rmIm S_{12}$, ...,$\rmRe S_{nn}$, $\rmIm S_{nn}$; $S$ is restricted to 
move on the surface determined by unitarity ($SS^{\dagger } \smeq I$) and, 
for $\beta  \smeq 1$, symmetry ($S \smeq S^{T}$).

We assume $E$ is far from thresholds and recall that again $S(E)$ is {\it %
analytic} in the upper half of the complex-energy plane. The study of the
statistical properties of $S$ is again simplified by idealizing $S(E)$, for
real $E$, as a {\it stationary random-matrix function} satisfying the
condition of {\it ergodicity}. The same argument as in the 1D case above
shows that the AE requirement (\ref{AE 1 ch.}) is generalized to
\begin{equation}
\left\langle \left( S_{a_{1}b_{1}}\right) ^{n_{1}}\cdot \cdot \cdot \left(
S_{a_{k}b_{k}}\right) ^{n_{k}}\right\rangle =\left\langle
S_{a_{1}b_{1}}\right\rangle ^{n_{1}}\cdot \cdot \cdot \left\langle
S_{a_{k}b_{k}}\right\rangle ^{n_{k}}.  \label{AE n ch.}
\end{equation}
Notice that this expression involves only $S$, and not $S^{*}$ matrix
elements. Similarly, if $f(S)$ is a function that can be expanded as a
series of nonnegative powers of $S_{11},\cdot \cdot \cdot ,S_{nn}$ (analytic
in $S$), we must have the reproducing property (\ref{repr. prop.}).

Our starting point is the invariant measure $d\mu _{\beta }(S)$ that was
introduced in the last subsection. The average of $S$ evaluated with that
measure vanishes (shown explicitly in section \ref{avg products S}), 
so that the prompt, or direct, components described in the Introduction
vanish. It is
easy to check that the AE requirements (\ref{AE n ch.}) or, equivalently,
the reproducing property (\ref{repr. prop.}), is satisfied exactly for the
invariant measure. Ensembles that contain more information than the
invariant one are constructed by multiplying the latter by appropriate
functions of $S$. We relate the probability density $p_{\left\langle
S\right\rangle }^{(\beta )}(S)$ to the differential probability through

\begin{equation}
dP_{\left\langle S\right\rangle }^{(\beta )}(S)=p_{\left\langle
S\right\rangle }^{(\beta )}(S)d\mu _{\beta }(S)  \label{dP<S> n}
\end{equation}
and require the fulfillment of the AE conditions. It was shown in Ref. \cite
{mello-pereyra-seligman} that, for $n>1$, the AE conditions and reality of
the answer are not enough to determine the probability distribution
uniquely. However, it was shown that the probability density ($V_{\beta }$
is a normalization factor) 
\begin{equation}
p_{\left\langle S\right\rangle }(S)=V_{\beta }^{-1}\frac{[{\rm det}%
(I-\left\langle S\right\rangle \left\langle S\right\rangle ^{\dag
})]^{(\beta n+2-\beta )/2}}{|{\rm det}(I-S\left\langle S\right\rangle ^{\dag
})|^{\beta n+2-\beta }},  \label{poisson ensemble n}
\end{equation}
known again as Poisson's kernel, not only satisfies the AE requirements (\ref
{AE n ch.})\cite{hua}, but the {\it information} ${\cal I}$ associated with it 
\begin{equation}
{\cal I}[p]\equiv \int p_{\langle S \rangle}(S)
\ln p_{\langle S \rangle}(S)d\mu (S)  \label{I[p]}
\end{equation}
is {\it less than or equal to that of any other probability density
satisfying the AE requirements for the same }$\left\langle S\right\rangle$
\cite{mello-pereyra-seligman}.  Notice that, for $n \smeq 1$, Eq.
(\ref{poisson ensemble n}) reduces to (\ref {poisson ensemble n=1}).  Thus
the information entering Poisson's kernel specifies : {\bf 1)} General
properties: {\it i)} {\it flux conservation} (giving rise to unitarity of
the $S$ matrix), {\it ii)} {\it causality} and the related analytical
properties of $S(E)$ in the complex-energy plane, and {\it iii) }the
presence or absence of time-reversal (and spin-rotation symmetry when spin
is taken into account), that determines the {\it universality class}:
orthogonal, unitary (or symplectic) {\bf 2)} A specific property: the
ensemble average $\left\langle S\right\rangle $ ($ \smeq  \overline{S}$
under ergodicity), which controls the presence of {\it prompt}, or {\it
direct processes} in the scattering problem.  System-specific {\it details
other than the optical S are assumed to be irrelevant.}

The fact that for $n>1$ AE and reality do not fix the ensemble uniquely is
not surprising. In general (the $1\times 1$ case being exceptional) we
expect, physically, the matrix $\left\langle S\right\rangle $ to be
insufficient to characterize the full distribution when, in addition to
the prompt and equilibrated components, there are other contributions
associated with different time scales \cite{agassi-et-al}. Out of all
possibilities, though, the information-theoretic argument selects the one
where the prompt and equilibrated components and the associated optical
$S$ are the only physically relevant quantities.

In addition to the completely general derivation of Poisson's kernel above,
we present a concrete construction of this distribution following Ref.
\cite{piet1,piet}. For the equilibrated part of the response, suppose there is
an $S$-matrix $S_0$ which is distributed according to the circular ensembles.
For the prompt response, imagine a scattering process $S_1$ occuring prior to
the response $S_0$. The total scattering is the composition of these two
parts. Specifically, imagine bunching the $L$ leads of the cavity 
into a ``superlead'' containing $n$ incoming and $n$ outgoing waves. 
Along the superlead, between the cavity and infinity, we
connect a scatterer of the appropriate symmetry class described by $S_1$. 
Since there are $n$ incoming and $n$ outgoing waves on either side of the 
scatterer, $S_1$ is $2n$-dimensional and can be written
\begin{equation}
S_{1}=\left[ 
\begin{array}{cc}
r_{1} & t_{1}^{\prime } \\ 
t_{1} & r_{1}^{\prime }
\end{array}
\right] .  \label{S1}
\end{equation}
The composition of the two scattering processes yields a total $S$ 
\begin{equation}
S = r_1 + t_{1}^{\prime} (1- S_0 r_{1}^{\prime})^{-1} S_0 t_{1}.
\label{S0}
\end{equation}
One can prove \cite{hua,piet,piet1,friedman-mello} the following 
statement:
the distribution of $S$ is Poisson's measure (\ref{poisson ensemble n}) 
with $\langle S \rangle  \smeq  r_{1}$ if and only if the distribution of
$S_0$ is the invariant measure.  That is, Eq. (\ref{S0}) transforms between 
the problem with direct processes and the one without (for the one-energy
distribution considered here). Also, one can show \cite{hua,friedman-mello} 
that the distribution is independent of the choice of $t_{1}$ and 
$t_{1}^{\prime }$, as long as they belong to a unitary matrix $S_{1}$.

Note that throughout this work we use arguments which refer only to physical
information expressible entirely in terms of the $S$-matrix. An alternate
point of view is to express everything in terms of an underlying Hamiltonian
which one then analyzes using statistical or information-theoretic assumptions.
These two points of view give, in fact, the same results:
one can prove \cite{hans,piet1,piet} that, for $\langle S \rangle  \smeq 0$, a 
Gaussian Ensemble for the underlying Hamiltonian gives a Circular Ensemble 
for the resulting $S$. The argument was extended to 
$ \langle S \rangle \smneq 0$ in Refs. \cite{piet1,piet} using the 
transformation (\ref{S0}) above.

\section{Absence of direct processes}
\label{no direct processes}

In this case the optical matrix $ \langle S \rangle $ vanishes and
Poisson's kernel (\ref{poisson ensemble n}) reduces to the invariant 
measure: 
\begin{equation}
\label{dP=dmu}
dP_{\left\langle S\right\rangle =0}^{(\beta )}= d\mu ^{(\beta)}(S) .
\end{equation}
We now derive the implications of this distribution for $T$, starting 
with the average and variance of $T$ and then turning to its distribution.

\subsection{Averages of products of $S$: Weak-localization and conductance 
fluctuations}
\label{avg products S}

Averages over the invariant measure of products of $S$ matrix elements
(invariant integration) can be evaluated using solely the properties of the
measure, without performing any integration explicitly \cite
{mello-seligman80,mello90}. We first discuss the unitary case, because it is
simpler, and then the orthogonal one.

\subsubsection{The case $\beta \smeq 2$}

To illustrate the procedure, consider first the average over the invariant
measure of a single $S$ matrix element, to be denoted as 
\begin{equation}
 \langle S_{a\alpha } \rangle _{0}^{(2)} =
\int S_{a\alpha }d\mu ^{(2)}(S) .
\label{<S>}
\end{equation}
Even though it is trivial to recognize that this average vanishes-- since the
invariant measure gives the same weight to $S_{a\alpha }$ and to its
negative-- we present a more formal argument, which will be generalized
later to more complicated averages. If $U^{0}$ is an {\it arbitrary} but 
{\it fixed} unitary matrix, we define the transformed $\widetilde{S}$ as 
\begin{equation}
S=U^{0}\widetilde{S}.  \label{S transformed}
\end{equation}
Introducing (\ref{S transformed}) in (\ref{<S>}) we have 
\begin{equation}
 {\langle S_{a\alpha } \rangle}_{0}^{(2)} =
\sum_{a^{\prime }}U_{aa^{\prime}}^{0}
\int \widetilde{S}_{a^{\prime }\alpha }d\mu^{(2)} (\widetilde{S})
=\sum_{a^{\prime }}U_{aa^{\prime}}^{0}
{\langle S_{a^{\prime }\alpha } \rangle}_{0}^{(2)}
\label{<S> 1}
\end{equation}
where we have used the defining property (\ref{inv measure}) of the invariant 
measure and the definition of 
$\left\langle S_{a^{\prime }\alpha }\right\rangle _{0}$.
In particular, if we take, as the arbitrary fixed matrix $U^{0}$%
\begin{equation}
U_{aa^{\prime }}^{0}=e^{i\theta _{a}}\delta _{aa^{\prime }},  \label{U0}
\end{equation}
we find 
\begin{equation}
\left\langle S_{a\alpha }\right\rangle _{0}=e^{i\theta _{a}}\left\langle
S_{a\alpha }\right\rangle _{0}.  \label{<S> 3}
\end{equation}
Since this expression should hold for arbitrary $\theta _{a}$, we conclude
that 
\begin{equation}
\left\langle S_{a\alpha }\right\rangle _{0}=0.  \label{<S> 4}
\end{equation}

The above argument can be generalized to prove that 
\begin{equation}
\left\langle 
\left[ S_{b_{1}\beta _{1}}\cdot \cdot \cdot S_{b_{p}\beta_{p}}\right] 
\left[ S_{a_{1}\alpha _{1}}\cdot \cdot \cdot S_{a_{q}\alpha_{q}}\right] 
^{^{*}}\right\rangle _{0}^{(2)} =0,  \label{<SpSstarq>}
\end{equation}
unless $p \smeq q$ and unless $\{a_{1}, \ldots, a_{p}\}$ and 
$\{b_{1}, \ldots, b_{p}\}$ 
constitute the same set of indices except for order, 
with the same condition for the sets
$\{\alpha _{1}, \ldots ,\alpha _{p}\}$, 
$\{\beta _{1}, \ldots ,\beta _{p}\}$.
In particular, consider $p \smeq q \smeq 1$. Using the same argument
as above, we find 
\begin{equation}
{\langle S_{b\beta }S_{a\alpha }^{^{*}} \rangle}_{0}^{(2)}
=\sum_{a^{\prime }b^{\prime }}U_{bb^{\prime }}^{0}\left( U_{aa^{\prime
}}^{0}\right) ^{*} {\langle S_{b^{\prime }\beta }S_{a^{\prime }\alpha
}^{^{*}} \rangle}_{0}^{(2)},  
\;\; \forall U^{0}. 
\label{<SSstar>}
\end{equation}
For $U^{0}$ given as in Eq. (\ref{U0}), we have 
\begin{equation}
\left\langle S_{b\beta }S_{a\alpha }^{^{*}}\right\rangle _{0}=e^{i(\theta
_{b}-\theta _{a})}\left\langle S_{b\beta }S_{a\alpha
}^{^{*}}\right\rangle _{0}  \label{<SSstar> 1}
\end{equation}
which vanishes unless $b \smeq a.$ Had we defined $\widetilde{S}$ through 
right multiplication instead of left multiplication in (\ref{S transformed}), 
we would have concluded $\beta \smeq \alpha $. The only
nonvanishing possibility in Eq. (\ref{<SSstar>}) is thus 
\begin{equation}
\left\langle \left| S_{a\alpha }\right| ^{2}\right\rangle
_{0}=\sum_{a^{\prime }}\left| U_{aa^{\prime }}^{0}\right| ^{2}\left\langle
\left| S_{a^{\prime }\alpha }\right| ^{2}\right\rangle _{0},  
\;\; \forall U^{0}. 
\label{<|S|^2>}
\end{equation}
For instance, the choice of the matrix that produces a permutation of the 
indices $1$ and $2$ as $U^{0}$ yields: 
\begin{equation}
U^{0}=\left[ 
\begin{array}{ccccc}
0 & 1 & 0 & \cdots & 0 \\ 
1 & 0 & 0 & \cdots & 0 \\ 
0 & 0 & 1 & \cdots & 0 \\ 
\vdots & \vdots & \vdots & \ddots & \vdots \\ 
0 & 0 & 0 & \cdots & 1
\end{array}
\right], \qquad  
\left\langle \left| S_{1\alpha }\right| ^{2}\right\rangle =\left\langle
\left| S_{2\alpha }\right| ^{2}\right\rangle .
\label{<|S1alpha|^2>=<|S2alpha|^2>}
\end{equation}
Similarly 
\begin{equation}
\left\langle \left| S_{1\alpha }\right| ^{2}\right\rangle _{0}=\cdots 
=\left\langle \left| S_{n\alpha }\right| ^{2}\right\rangle _{0}.
\label{<|S1alpha|^2>=<|Snalpha|^2>}
\end{equation}
The final result for the average intensity follows from unitarity
\begin{equation}
\sum_{a=1}^{n}\left\langle \left| S_{a\alpha }\right| ^{2}\right\rangle
_{0}=1
\quad \Longrightarrow \quad
{\left\langle \left| S_{a\alpha }\right| ^{2}\right\rangle}_{0}^{(2)}
=\frac{1}{n}.  
\label{<|Saalpha|^2>}
\end{equation}

As an application, we calculate the {\it average conductance} when we have a
cavity connected to the outside by means of two leads supporting $N_{1}$ and 
$N_{2}$ open channels (so that $n \smeq $ $N_{1}+N_{2})$: 
\begin{equation}
\left\langle T\right\rangle _{0}^{(2)}=
\sum_{a=1}^{N_{1}}\sum_{b=1}^{N_{2}}\left\langle \left| t_{ab}\right|
^{2}\right\rangle _{0}^{(2)} 
=\frac{N_{1}N_{2}}{N_{1}+N_{2}}
= {\left[ \frac{1}{N_{1}}+\frac{1}{N_{2}} \right]}^{-1} ,  
\label{<g> beta=2}
\end{equation}
which is the series addition of the two conductances $N_{1}$ and $N_{2}$.
(The superscript on the angular brackets indicates $\beta \smeq 2$.)

With similar arguments one finds \cite{mello90}: 
\begin{eqnarray}
\left\langle \left| S_{12}\right| ^{2}\left| S_{34}\right| ^{2}\right\rangle
_{0}^{(2)}=\frac{1}{n^{2}-1} \; ,
\label{<|S12|^2|S34|^2> beta=2}
\\[0.1in]
\left\langle \left| S_{12}\right| ^{2}\left| S_{13}\right| ^{2}\right\rangle
_{0}^{(2)}=\frac{1}{n(n+1)} \; ,  
\label{<|S12|^2|S13|^2> beta=2}
\\[0.1in]
\left\langle \left| S_{12}\right| ^{4}\right\rangle _{0}^{(2)}=\frac{2}{%
n(n+1)}.  
\label{<|S12|^4> beta=2}
\end{eqnarray}
Here, 1,2,3,4 stand for any quartet of different indices, so that $n$ in 
each case must be large enough to accomadate as many indices as necessary.

As an application, we calculate the second moment of the conductance as 
\begin{equation}
\left\langle T^{2}\right\rangle _{0}^{(2)}=
 \sum_{a,c=1}^{N_1} \sum_{b,d=1}^{N_2} \left\langle
\left| t_{ab}\right| ^{2}\left| t_{cd}\right| ^{2}\right\rangle _{0}^{(2)} 
=\frac{N_{1}^{2}N_{2}^{2}}{\left( N_{1}+N_{2}\right) ^{2}-1}.
\label{<g^2> beta=2}
\end{equation}
The variance of the conductance is then given by 
\begin{equation}
\left[ \rmvar(T)\right] _{0}^{(2)}=\frac{N_{1}^{2}N_{2}^{2}}{%
(N_{1}+N_{2})^{2}\left[ \left( N_{1}+N_{2}\right) ^{2}-1\right] }.
\label{var(g) beta=2}
\end{equation}
Notice that in the limit $N_{1}$, $N_{2}\rightarrow \infty $ 
with $N_{1}/N_{2} \smeq K$ fixed
\begin{equation}
[ \rmvar(T)]_0^{(2)} \rightarrow \frac{K^{2}}{\left( K+1\right) ^{4}},
\label{var(g) K beta=2}
\end{equation}
a constant which depends only on the ratio $N_{1}/N_{2}$ and on no other
details of the cavity. 
Since this limit corresponds to increasing the width of the waveguides
and so of the full system, the fact that the result is constant is the
analog of the so-called {\it universal conductance fluctuations (UCF)} 
well-known for quasi-1D disordered systems \cite{AltLeeWebb91,BvH91}. 
In particular, for $K \smeq 1$, $\rmvar (T)\rightarrow 1/16$,
slightly less than the quasi-1D value of $1/15$.

\subsubsection{The case $\beta \smeq 1$}

Just as above, we first illustrate the procedure through the average over
the invariant measure of a single $S$ matrix element
[Eq. (\ref{<S>}) written for $\beta  \smeq 1$]
which vanishes trivially. We introduce the transformed $\widetilde{S}$, again
a unitary symmetric matrix, through 
\begin{equation}
S=U^{0}\widetilde{S}(U^{0})^{T}.  \label{S tilde 1}
\end{equation}
Substituting (\ref{S tilde 1}) in 
the integral definition of 
$\left\langle S_{ab}\right\rangle _{0}^{(1)}$ 
and using the defining
property of the invariant measure (\ref{inv measure}) we have 
\begin{equation}
\left\langle S_{ab}\right\rangle _{0}^{(1)} =\sum_{a^{\prime }}U_{aa^{\prime
}}^{0}U_{bb^{\prime }}^{0}\int \widetilde{S}_{a^{\prime }b^{\prime }}d\mu
^{(1)}(\widetilde{S})
=\sum_{a^{\prime }}U_{aa^{\prime
}}^{0}U_{bb^{\prime }}^{0}\left\langle S_{a^{\prime }b^{\prime
}}\right\rangle _{0}^{(1)} .  
\label{<S> 6}
\end{equation}
In particular, if we take as the arbitrary fixed matrix $U^{0}$ the one
given by Eq. (\ref{U0}), we find 
\begin{equation}
\left\langle S_{ab}\right\rangle _{0}=e^{i(\theta _{a}+\theta
_{b})}\left\langle S_{ab}\right\rangle _{0}
\quad \Longrightarrow \quad
\left\langle S_{ab}\right\rangle _{0}=0
\label{<S> 8}
\end{equation}
since $\theta _{a}$, $\theta _{b}$ are arbitrary.

Just as for $\beta  \smeq 2$, the above argument can be generalized to 
prove that 
\begin{equation}
\left\langle \left[ S_{a_{1}b_{1}}\cdot \cdot \cdot S_{_{a_{p}b_{p}}}\right]
\left[ S_{\alpha _{1}\beta _{1}}\cdot \cdot \cdot S_{\alpha _{q}\beta
_{q}}\right] ^{^{*}}\right\rangle _{0}^{(1)} =0,  \label{<SpSstarq > 1}
\end{equation}
unless $p \smeq q$ and unless $\{ a_{1}, b_{1}, \ldots, a_{p}, b_{p}\}$ and 
$\{\alpha _{1}, \beta _{1}, \ldots ,\alpha _{p}, \beta _{p}\}$ 
constitute the same set of indices except for order.
In particular, for $p \smeq q \smeq 1$ we find 
\begin{equation}
\left\langle S_{ab}S_{\alpha \beta }^{^{*}}\right\rangle
_{0}^{(1)} =\sum_{a^{\prime }b^{\prime }\alpha ^{\prime }\beta ^{\prime
}}U_{aa^{\prime }}^{0}U_{bb^{\prime }}^{0}\left( U_{\alpha \alpha ^{\prime
}}^{0}U_{\beta \beta ^{\prime }}^{0}\right) ^{^{*}}\left\langle S_{a^{\prime
}b^{\prime }}S_{\alpha ^{\prime }\beta ^{\prime }}^{^{*}}
\right\rangle _{0}^{(1)},
\; \forall U^{0}. 
\end{equation}
For instance, for $U^{0}$ given as in Eq. (\ref{U0}), we have 
\begin{equation}
\left\langle S_{ab}S_{\alpha \beta }^{^{*}}\right\rangle _{0}^{(1)}
=e^{i(\theta
_{a}+\theta _{b}-\theta _{\alpha }-\theta _{\beta })}\left\langle
S_{ab}S_{\alpha \beta }^{^{*}}\right\rangle _{0}^{(1)} ,  
\label{<S Sstar> 1}
\end{equation}
which vanishes unless 
$\{a,b\} \smeq \{\alpha ,\beta \}$ or $\{\beta ,\alpha \}$.
As a particular case, take $a \smeq b \smeq \alpha \smeq \beta \smeq 1$ and, 
as $U^{0}$, a matrix with the structure 
\begin{equation}
U^{0}=\left[ 
\begin{array}{ccccc}
U_{11}^{0} & U_{12}^{0} & 0 & \cdots & 0 \\ 
U_{21}^{0} & U_{22}^{0} & 0 & \cdots & 0 \\ 
0 & 0 & 1 & \cdots & 0 \\ 
\vdots & \vdots & \vdots & \ddots & \cdots \\ 
0 & 0 & 0 & \cdots & 1
\end{array}
\right] .  \label{U0 2}
\end{equation}
We find 
\begin{eqnarray}
\fl
\left\langle \left| S_{11}\right| ^{2}\right\rangle _{0}=\sum_{a^{\prime
}b^{\prime }}U_{1a^{\prime }}^{0}U_{1b^{\prime }}^{0}\left( U_{1a^{\prime
}}^{0}U_{1b^{\prime }}^{0}\right) ^{^{*}}\left\langle S_{a^{\prime
}b^{\prime }}S_{a^{\prime }b^{\prime }}^{^{*}}\right\rangle _{0} 
\nonumber \\[0.05in]
\lo+
\sum_{a^{\prime }\neq b^{\prime }}U_{1a^{\prime }}^{0}U_{1b^{\prime
}}^{0}\left( U_{1b^{\prime }}^{0}U_{1a^{\prime }}^{0}\right)
^{^{*}}\left\langle S_{a^{\prime }b^{\prime }}S_{b^{\prime }a^{\prime
}}^{^{*}}\right\rangle _{0} 
\nonumber \\[0.05in]
\lo=
\left[ \left| U_{11}^{0}\right| ^{4}+\left| U_{12}^{0}\right| ^{4}\right]
\left\langle \left| S_{11}\right| ^{2}\right\rangle _{0}+4\left|
U_{11}^{0}\right| ^{2}\left| U_{12}^{0}\right| ^{2}\left\langle \left|
S_{12}\right| ^{2}\right\rangle _{0}.  
\label{<S11S11star>}
\end{eqnarray}
Squaring the unitarity relation 
$ \left| U_{11}^{0}\right| ^{2}+\left| U_{12}^{0}\right| ^{2} \smeq 1 $
and substituting in Eq. (\ref{<S11S11star>}) we finally obtain 
\begin{equation}
\left\langle \left| S_{11}\right| ^{2}\right\rangle _{0}^{(1)}
=2\left\langle
\left| S_{12}\right| ^{2}\right\rangle _{0}^{(1)}.  
\label{enhancement}
\end{equation}
This result is very important. It states that time-reversal invariance
(TRI)
has the consequence that {\it the average of the absolute value squared
of a diagonal }$S${\it -matrix element is twice as large as that of an
off-diagonal one}, under the invariant measure. By unitarity, the specific 
value of each one of these averages is given by 
\begin{equation}
\left\langle \left| S_{aa}\right| ^{2}\right\rangle _{0}^{(1)}
=\frac{2}{n+1} , \qquad
\left\langle \left| S_{a\neq b}\right| ^{2}\right\rangle _{0}^{(1)}
=\frac{1}{n+1}.
\label{<SabSab star>}
\end{equation}

Just as in the above case $\beta  \smeq 2$, we calculate as an application the 
{\it average conductance} when our cavity is connected to the outside by two
leads with $N_{1}$ and $N_{2}$ open channels ($n \smeq $ $N_{1}+N_{2})$: 
\begin{equation}
\left\langle T\right\rangle
_{0}^{(1)}=\sum_{a=1}^{N_{1}}\sum_{b=1}^{N_{2}}\left\langle \left|
t_{ab}\right| ^{2}\right\rangle _{0}^{(1)} 
=\frac{N_{1}N_{2}}{N_{1}+N_{2}+1}.  \label{<g> beta=1}
\end{equation}
Here, the extra $1$ in the denominator as compared with Eq. (\ref{<g>
beta=2}) is the {\it weak-localization correction} (WLC), a symmetry effect
resulting from TRI. We can rewrite Eq. (\ref{<g> beta=1}) separating
out the WLC term as 
\begin{equation}
\left\langle T\right\rangle _{0}^{(1)}=\frac{N_{1}N_{2}}{N_{1}+N_{2}}-\frac{%
N_{1}N_{2}}{\left( N_{1}+N_{2}\right) \left( N_{1}+N_{2}+1\right) }.
\label{<g> beta=1 1}
\end{equation}
In particular, for $N_{1} \smeq N_{2} \smeq N$ and for $N\rightarrow \infty $, 
corresponding to a large system, the WLC term tends to the universal 
number $-1/4$.

Using a similar procedure, one finds the results ($n$ being again the
dimensionality of the $S$ matrix)\cite{mello-seligman80} 
\begin{eqnarray}
\left\langle \left| S_{12}\right| ^{2}\left| S_{34}\right| ^{2}\right\rangle
_{0}^{(1)}=\frac{n+2}{n(n+1)(n+3)} \; , 
\label{<|S12|^2|S34|^2> beta=1}
\\[0.1in]
\left\langle \left| S_{12}\right| ^{2}\left| S_{13}\right| ^{2}\right\rangle
_{0}^{(1)}=\frac{1}{n(n+3)} \; ,
\label{<|S12|^2|S13|^2> beta=1}
\\[0.1in]
\left\langle \left| S_{12}\right| ^{4}\right\rangle _{0}^{(1)}=\frac{2}{%
n(n+3)}.  
\label{<|S12|^4> beta=1}
\end{eqnarray}
A comment like that made right after Eq. (\ref{<|S12|^4> beta=2}) 
applies here as well.
Just as in Eq. (\ref{<g^2> beta=2}), we now find for the second moment of
the conductance 
\begin{equation}
\left\langle T^{2}\right\rangle _{0}^{(1)}=\frac{N_{1}N_{2}\left[
N_{1}N_{2}\left( N_{1}+N_{2}+2\right) +2\right] }{%
(N_{1}+N_{2})(N_{1}+N_{2}+1)(N_{1}+N_{2}+3)}  \label{<g^2> beta=1}
\end{equation}
and for its variance 
\begin{equation}
\left[ \rmvar(T)\right] _{0}^{(1)}=\frac{2N_{1}N_{2}\left( N_{1}+1\right)
\left( N_{2}+1\right) }{(N_{1}+N_{2})\left( N_{1}+N_{2}+1\right) ^{2}\left(
N_{1}+N_{2}+3\right) }  
\rightarrow
2\frac{K^{2}}{\left( K+1\right) ^{4}}
\label{var(g) beta=1}
\end{equation}
where in the limit $N_{1}$, $N_{2}\rightarrow \infty $ with $N_{1}/N_{2}
\smeq K$, a fixed number.  Note that in this universal limit, the variance
here is exactly twice as large as for $\beta  \smeq 2$, Eq. (\ref{var(g) K
beta=2}), another result of time-reversal invariance.  For the particular
case $K \smeq 1$, the limiting value of the variance is $1/8$.

\subsection{The distribution of the conductance in the two-equal-lead case}
\label{w(T)}

In the last section we characterized the conductance of a cavity through
the first two moments of its transmission. If the distribution of the
conductance is Gaussian, this is sufficient to characterize the full
distribution.  In fact, it can be shown that in the large-size universal
limit, $N \rightarrow \infty$, the distribution of the conductance is
indeed Gaussian \cite{Politzer}.

In general, however, the probability density of $T$ will not be Gaussian,
and it is of interest, then, to derive results for this density. For this
purpose, the polar representation of Sec. \ref{parametrization S} is 
particularly
useful since the conductance is directly related to the $\{ \tau \}$ whose
joint probability distribution we know. Specifically,
the distribution of the transmission $T$ of Eq. (\ref{g(tau)}) can be
obtained by direct integration of the $P^{(\beta )}(\{\tau \})$ of 
Eq. (\ref{w(tau)}):
\begin{equation}
w^{(\beta )}(T) = \int \delta (T- \sum _a \tau _a )
P^{(\beta )} \left( \{ \tau _a \}\right) \prod _a d\tau _a  .
\label{def of w(T)}
\end{equation}
We consider a few examples below.

\subsubsection{The case $N \smeq 1$}

In this case we have only one $\tau _a$, that we may call $\tau $, so that 
$T \smeq \tau $, and $0\leq T\leq 1$. Eq. (\ref{w(tau)}) then gives
\begin{equation}
w^{(1)}(T)=\frac{1}{2\sqrt{T}}, \qquad  w^{({2})}(T)=1 .  
\label{w(g) N=1}
\end{equation}
For $\beta  \smeq 1$,
we thus have a higher probability to find small $T$'s than $T \sim 1$: this
is clearly a symmetry effect, a result of TRI that favors backscattering
and hence low conductances.

\subsubsection{The case $N \smeq 2$}

Now $T \smeq \tau _{1}+\tau _{2}$, and $0\leq T\leq 2$. In 
\ref{w(T) no direct} we show that 
\begin{equation}
w^{({1})}(T)= 
\left\{
\begin{array}{cl}
\frac{3}{2}T, & 0<T<1 \\ 
\\
\frac{3}{2}\left( T-2\sqrt{T-1}\right) ,  & 1<T<2
\end{array}
\right.
\label{w(g) N=2 beta=1}
\end{equation}
and
\begin{equation}
w^{({2})}(T)=2\left[ 1-\left| 1-T\right| \right] ^{3}
\label{w(g) N=2 beta=2}
\end{equation}
For $\beta  \smeq 1$, notice the square-root cusp at $T \smeq 1$. We find,
once again, a higher probability for $T<1$ than for $T>1$ to occur. On the
other hand, for $\beta  \smeq 2$, $w(T)$ is again symmetric around $T
\smeq 1$.

\subsubsection{The case $N \smeq 3$}

Now $T \smeq \tau _1+\tau _2+\tau _3$, and $0\leq T\leq 3$. 
For $\beta  \smeq 2$ one finds
\begin{equation}
w^{({2})}(T)\!=
\!\left\{
\begin{array}{cl}
\frac{9}{42}T^8, & 0\!<T\!<1 \\
\\
- \frac{2781}{14} + \frac{6588}{7}T - {\scriptstyle 1818}T^2 + 
{\scriptstyle 1836}T^3 
& \\
- {\scriptstyle 1035}T^4 + {\scriptstyle 324}T^5 - {\scriptstyle 54}T^6 
+ \frac{36}{7} T^7 - \frac{3}{7} T^8 , & 1\!<T\!<\frac{3}{2} \\
\end{array}
\right.
\end{equation}
and the distribution is symmetric about $T \smeq 1.5$.  As mentioned above,  
$w^{({\beta })}(T)$ gradually approaches a Gaussian distribution. 

\subsubsection{Arbitrary $N$}

In this case it is straightforward to obtain the dependence of the tail
of the distribution in the region $0 < T < 1$. In this region the constraint
that $\tau < 1$ does not enter; a calculation given in \ref{w(T) no direct}
shows that
\begin{equation}
w_N^{(\beta )} (T) \propto T^{\beta N^2/2 -1}.
\label{w(g) N arb}
\end{equation}

\section{Presence of direct processes}
\label{direct processes}

In order to treat cases involving direct processes, as in billiards in
which short paths produce a prompt response, we now need Poisson's kernel
(\ref{poisson ensemble n}) in its full generality. We discuss below some
analytical results for the distribution of the spinless conductance $T$ in
the case of a cavity connected to the outside by means of two leads
supporting one open channel each ($N_{1} \smeq N_{2} \smeq 1)$, giving
rise to a 2-dimensional $S$ matrix. There is only one $\tau $ in this case
and it is its distribution that we seek, since $T \smeq \tau $. While the
expressions that we derive are somewhat cumbersome, they are used for
comparing to numerical results in section \ref{numerics} where plots of
several examples are displayed.

We write the optical $S$-matrix $\overline{S}$, a subunitary matrix, as 
\begin{equation}
\overline{S}=\left[ 
\begin{array}{cc}
x & w \\ 
z & y
\end{array}
\right] ,  \label{form of <S>}
\end{equation}
where the entries are, in general, complex numbers.

\subsection{The case $\beta  \smeq 2$}

From Eq. (\ref{poisson ensemble n}) we write the differential probability
for the $S$ matrix as
\begin{equation}
dP_{\overline{S}}^{(2)}(S)=\frac{[{\rm det}(I-\overline{S}\,
\overline{S}^{\dag}
)]^{n}}{|{\rm det}(I-S\,\overline{S}^{\dag })|^{2n}}d\mu _{0}^{(2)}(S),
\label{dP(S)}
\end{equation}
where 
\begin{equation}
d\mu _{0}^{(2)}(S)=\frac{d\mu ^{(2)}(S)}{V},\qquad \int d\mu _{0}^{(2)}(S)=1.
\label{dmu0}
\end{equation}
We are interested here in the case $n \smeq 2$.

In the polar representation (\ref{S polar repr}) with $n \smeq 2$, 
the $S$ matrix has the form 
\begin{equation}
S=\left[ 
\begin{array}{cc}
e^{i\alpha } & 0 \\ 
0 & e^{i\beta }
\end{array}
\right] \left[ 
\begin{array}{cc}
-\sqrt{1-\tau } & \sqrt{\tau } \\ 
\sqrt{\tau } & \sqrt{1-\tau }
\end{array}
\right] \left[ 
\begin{array}{cc}
e^{i\gamma } & 0 \\ 
0 & e^{i\delta }
\end{array}
\right] ,  \label{S(alpha,...)}
\end{equation}
the invariant measure (\ref{dmu0}) being [see Eqs. (\ref{inv meas (tau v)}),
(\ref{w(tau)})] 
\begin{equation}
d\mu _{0}(S)=d\tau \frac{d\alpha d\beta d\gamma d\delta }{\left( 2\pi
\right) ^{4}}.  \label{dmu0 1}
\end{equation}

1) As a particular case, suppose the optical $S$ matrix is diagonal, so that
there is {\it only direct reflection} and no direct transmission: in
Eq. (\ref{form of <S>}) we choose $w \smeq z \smeq 0$.

Substituting Eqs. (\ref{form of <S>}), (\ref{S(alpha,...)}), (\ref{dmu0 1})
in (\ref{dP(S)}) we find 
\begin{equation}
dP_{X,Y}^{(2)}(\tau ,\varphi ,\psi )
=\frac{\left( 1-X^{2}\right) ^{2}\left( 1-Y^{2}\right) ^{2}}{\left| \left(
e^{-i\varphi }+X\sqrt{1-\tau }\right) \left( e^{-i\psi }-Y\sqrt{1-\tau }%
\right) -XY\tau \right| ^{4}}\frac{d\tau d\varphi d\psi }{\left( 2\pi
\right) ^{2}}  \label{dP(tau,phi,psi)}
\end{equation}
where $\varphi \smeq \alpha +\gamma $, $\psi  \smeq \beta +\delta $, 
$X \smeq \left| x\right| 
$, $Y \smeq \left| y\right| $. The distribution of the conductance $T$ is thus 
\begin{eqnarray}
w_{X,Y}^{(2)}(T) & \smeq \left( 1-X^{2}\right) ^{2}\left( 1-Y^{2}\right) ^{2}
\\
& \times \left\langle \frac{1}{\left| \left( e^{-i\varphi }+X\sqrt{1-T}\right)
\left( e^{-i\psi }-Y\sqrt{1-T}\right) -XYT\right| ^{4}}\right\rangle
_{\varphi ,\psi }  \nonumber
\label{w(T) X,Y beta=2 a}
\end{eqnarray}
where $\left\langle \cdots \right\rangle _{\varphi ,\psi }$ denotes an average
over the variables $\varphi $, $\psi $. The result is ($0<T<1)$ 
\begin{equation}
w_{X,Y}^{(2)}(T)=K\frac{A-B(1-T)+C(1-T)^{2}+D(1-T)^{3}}
{\left[ E-2F(1-T)+G(1-T)^{2}\right] ^{5/2}},
\label{w(T) X,Y beta=2 b}
\end{equation}
where 
\[
K=(1-X^{2})^{2}(1-Y^{2})^{2}
, \qquad
A=(1-X^{4}Y^{4})(1-X^{2}Y^{2})
\]
\[
B=(X^{2}+Y^{2})(1-6X^{2}Y^{2}+X^{4}Y^{4})+4X^{2}Y^{2}(1+X^{2}Y^{2})
\]
\[
C=(1+X^{2}Y^{2})(6X^{2}Y^{2}-X^{4}-Y^{4})-4X^{2}Y^{2}(X^{2}+Y^{2})
\]
\[
D=(X^{2}+Y^{2})(X^{2}-Y^{2})^{2}
, \qquad
E=(1-X^{2}Y^{2})^{2}
\]
\[
F=(1+X^{2}Y^{2})(X^{2}+Y^{2})-4X^{2}Y^{2}
, \qquad
G=(X^{2}-Y^{2})^{2}.
\]

This result reduces to 1 when $X \smeq Y \smeq 0$, as it should. A particularly
interesting case is that of ``equivalent channels'', i.e. $X \smeq Y$, in which
the above expression reduces to
\begin{equation}
w_{X,X}^{(2)}(T)=(1-X^{2})\frac{(1-X^{4})^{2}+2X^{2}\left( 1+X^{4}\right)
T+4X^{4}T^{2}}{\left[ (1-X^{2})^{2}+4X^{2}T\right] ^{5/2}}.
\label{w(T) X=Y beta=2}
\end{equation}
The structure of this result is clear if we notice that
\[
w_{X,X}^{(2)}(0)=\left[ \frac{1+X^{2}}{1-X^{2}}\right] ^{2}>1
, \qquad
w_{X,X}^{(2)}(1)=\frac{\left( 1-X^{2}\right) \left( 1+X^{4}\right) }{\left(
1+X^{2}\right) ^{3}}<1,
\]
and hence $w_{X,X}^{(2)}(0)>w_{X,X}^{(2)}(1)$, so that small
conductances
are emphasized, as expected, because of the presence of direct reflection
and no direct transmission.

2) The case of {\it only direct transmission} and no direct reflection is
obtained by setting $x \smeq y \smeq 0$ in Eq.  (\ref{form of <S>}). The 
conductance distribution is obtained from Eq. (\ref{w(T) X,Y beta=2 b}) with 
the replacement $X\rightarrow W \smeq \left| w\right| $, 
Y$\rightarrow Z \smeq \left| z\right| $, $T\rightarrow 1-T$. 
In the equivalent-channel case we now obtain
a conductance distribution that emphasizes large conductances.

3) The case of a general optical $S$ matrix, Eq. (\ref{form of <S>}), has
also been worked out and the result expressed in terms of quadratures: 
because of its complexity, it will not be quoted here.

\subsection{The case $\beta  \smeq 1$}

This case is more complicated than that for $\beta  \smeq 2$ and we have only
succeeded in treating analytically some particular cases. Take, for instance,
a diagonal $\overline{S}$, i.e. $w \smeq z \smeq 0$ in Eq. (\ref{form of <S>}). 
With the same notation as above, we find 
\begin{eqnarray}
w_{X,Y}^{(1)}(T) & =\left| \left( 1-X^{2}\right) \left( 1-Y^{2}\right) \right|
^{3/2}\frac{1}{2\sqrt{T}}
\\
& \times \left\langle \frac{1}{\left| \left( e^{-i\varphi }+X\sqrt{1-T}\right)
\left( e^{-i\psi }-Y\sqrt{1-T}\right) -XYT\right| ^{3}}\right\rangle
_{\varphi ,\psi }, \nonumber  
\label{w(T) X,Y beta=1}
\end{eqnarray}
a result that has to be integrated numerically. When $X \smeq Y \smeq 0$, 
the distribution (\ref{w(T) X,Y beta=1}) reduces to $1/2\sqrt{T}$, as it should. 
It is interesting to notice that for $Y \smeq 0$ the above result can be 
integrated analytically, to give
\begin{equation}
w_{X,0}^{(1)}(T)=\frac{\left( 1-X^{2}\right) ^{3/2}}{2\sqrt{T}}%
\, _{2}F_{1}\left( 3/2,3/2;1;X^{2}(1-T)\right) ,
\end{equation}
$_{2}F_{1}$ being a hypergeometric function \cite{abramowitz-stegun}.

\section{Comparison with numerical calculations}
\label{numerics}

The information-theoretic approach that we have been discussing is expected
to be valid for cavities in which the classical dynamics is completely
chaotic, a property that refers to the {\it long time} behavior of the system. 
It is in such structures that the long time response
is ergodic and equilibrated, and so one can expect that maximum entropy
considerations will play a role. In this section we examine particular
cavities numerically in order to determine to what extent the 
information-theoretic approach really holds. The structures that we consider
are all ``billiards''-- they consist of hard walls surrounding a cavity
with constant potential-- with two leads. We start by considering 
particularly simple structures, then treat structures in which the absence of
direct processes is assured, then move on to structures having particularly
obvious direct processes.

\subsection{Simple structures}
\label{simple structures}

In the ``quantum chaos'' literature-- the study of how quantum properties 
depend on the nature of the classical dynamics in a system-- several
billiards are used as standard examples of closed chaotic systems. 
The two most studied are the Sinai billiard-- the region enclosed between a
square and a circle centered in the square-- and the stadium billiard--
two half-circles joined by straight edges. The classical dynamics in these
two billiards is known to be completely chaotic. For a test case open system,
then, it is natural to take one of these billiards and attach two leads.
The open stadium billiard was studied previously for this 
reason \cite{Doron91,JalBarSto90,Jensen91,BarJalSto93,BarJalSto93a}. 
Here we directly compare results for this system to the predictions of the 
information-theoretic approach.

The numerical methods used in these calculations are covered in detail in
Ref. \cite{Bar91}. Briefly, the procedure consists of the following three
steps.  First, discretize the Hamiltonian onto a square mesh using the
simplest finite-difference scheme. Solving the Schr\"odinger equation is
then reduced to solving a set of linear equations.  Second, find the Green
function at the desired energy from one lead to the other using a recursive
procedure and outgoing-wave boundary conditions.  This procedure essentially
uses the sparseness of the finite-difference matrix to efficiently solve the
linear equations.  Third, note that the transmission amplitude can be
obtained from this Green function by simply projecting onto the transverse
wave-functions in the lead.  The main parameter in these calculations, $ka$,
is the size of the mesh compared to the wavelength. In the results shown
here, $ka$ is always less than $0.8$ and $ka < 0.5$ in most cases. For these
values, the anisotropy of the Fermi surface is small; the non-parabolicity
of the dispersion is larger, but does not concern us here since we treat
transport at a fixed energy.

\begin{figure}[bt]
   \begin{center}
   \leavevmode
   \epsfxsize = 10.443cm
   \epsfbox{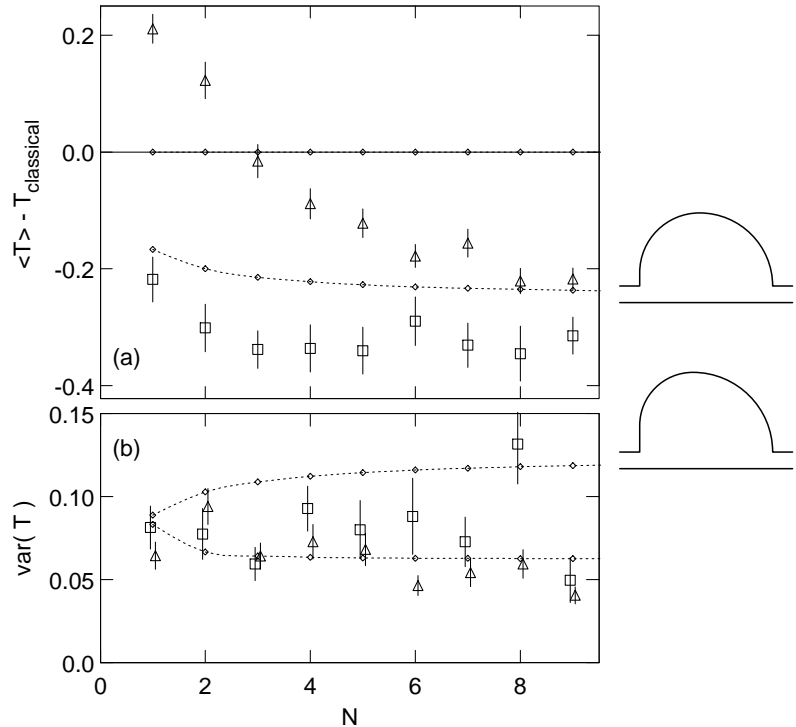}
   \end{center}
   \vspace*{-0.5cm}\caption{
   The magnitude of (a) the quantum correction to the classical conductance
   and (b) the conductance fluctuations as a function of the number of modes 
   in the lead $N$. 
   The two asymmetric stadium-like billiards shown at the right were
   used; the average of the results for the two cavities is shown. 
   The numerical results for $B \smeq 0$ (squares with statistical
   error bars)  and $B \smneq 0$ (triangles) are compared to the predictions 
   of the COE (points, dotted line) and CUE (points, dashed line).
   The agreement is poor.
   For both billiards, 25 energies were sampled for each value of $N$, and for 
   non-zero field $BA/\phi_0  \smeq  2,$ $4$ where $A$ is the area of the cavity. 
   }
   \label{num_simple_T}
\end{figure}

Two simple quantities to calculate both numerically and theoretically are
the average and variance of the conductance. The analytic results for the
invariant measure are in section \ref{avg products S}. Fig.
\ref{num_simple_T} shows the numerical results for the two asymmetric open
stadium shown on the side-- asymmetric half-stadiums are used in order to
avoid the complications of reflection symmetry.  The top panel shows the
deviation of the average transmission from the classical value of the
transmission. This classical value was obtained numerically by tracing
trajectories through the cavity. For fully equilibrated scattering, the
classical value is $N/2$ where $N$ is the number of channels in each lead, 
but in Fig. \ref{num_simple_T} $T_{\mbox{classical}}/N$ is $0.60$ ($0.58$)
for the upper (lower) cavity.  The bottom panel
shows the variance of $T$. While the numerical results are similar in
magnitude to the predictions, the agreement is clearly not very good.

\begin{figure}[tb]
   \begin{center}
   \leavevmode
   \epsfxsize = 0.95\textwidth
   \epsfbox{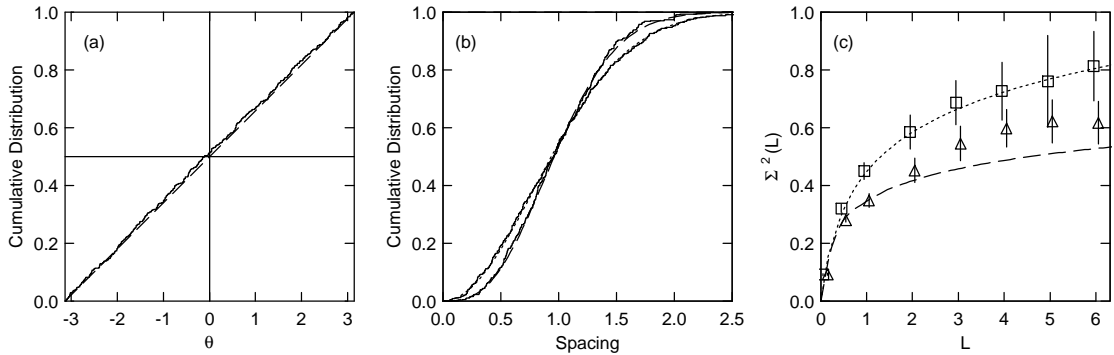}
   \end{center}
   \vspace*{-0.5cm}\caption{
   Statistics of the eigenphases of the $S$-matrix for the second stadium 
   shown in Fig. \protect\ref{num_simple_T} with $N \smeq 9$.
   (a) Cumulative distribution of the eigenphase density (solid line) 
   compared to the CE (dashed). 
   (b) Cumulative distribution of the difference between nearest-neighbor
   phases for both $B\smeq 0$ and $BA/\phi_0 \smeq 4$ compared to the 
   COE (dotted) and CUE (dashed). The spacing is normalized to the mean 
   separation.
   (c) Variance of number of phases in an interval $L$ for both $B\smeq 0$ 
   (squares) and $BA/\phi_0 \smeq 4$ (triangles) compared to the 
   COE (dotted) and CUE (dashed).
   All three statistics agree with the prediction of the circular
   ensembles-- constant density, a spacing distribution given
   by the Wigner surmise, and a logarithmically increasing
   variance-- despite the poor agreement for the transmission in
   Fig. \protect\ref{num_simple_T}.
   }
   \label{num_simple_theta}
\end{figure}

Before proceeding with our discussion of the conductance in these cavities,
we step back to perform the most common test of random matrix theory. In the
context of closed systems, it is natural to consider the statistics of the
energy levels. The degree to which these statistics agree with the
Wigner-Dyson statistics derived from random matrix theory is often used as
the prime indication of the validity of the theory for a given system. In
the context of the $S$-matrix, the analog is to look at the statistics of
the eigenphases: in the large $N$ limit their statistics is also
Wigner-Dyson \cite{mehta,porter}. In fact, studies of eigenphases of chaotic
scattering systems were carried out prior to any interest in the conductance
\cite{Gutz83,Blu88}. The statistics of the eigenphases can be characterized
by three representative quantities \cite{mehta,porter}.  First, the mean
density of eigenphases measures the uniformity of the system; for the
circular ensembles (CE) it is constant. Second, the nearest neighbor
distribution highlights the repulsion at short scales; it is approximately
the Wigner surmise for the CE. Third, the variance of the number of phases
within a certain range $L$, denoted $\Sigma^2 (L)$, indicates the rigidity
of the spectrum at large scales; it grows logarithmically with $L$ for the
CE.

These three quantities are shown for the simple open stadium in Fig.
\ref{num_simple_theta}.  The agreement with the predictions of the CE is
good for all three, in contrast to the results for the conductance above,
despite $N$ not being very large. Since the conductance involves the
transmission coefficient which is a property of the wavefunctions, this
indicates that the distribution of the eigenvectors is more sensitive to
deviations from the CE than the distribution of eigenphases. Thus the
eigenphase statistics cannot be taken as a definitive indication of the
validity of the CE: it is perfectly possible to have excellent eigenvalue
statistics while having poor eigenvector statistics
\cite{mucciolo-private}.

Returning to the properties of the conductance, we believe that the
deviations from the CE seen in the numerics (Fig. \ref{num_simple_T}) are
caused by the presence of short paths in these simple structures. This means
that the response is not fully equilibrated. The two most obvious types of
short paths in these structures are the direct paths between the leads and
the whispering gallery paths-- those that hit only the half-circle.  Short
paths will be included in the analysis in section \ref{numerics with d.p.}
below.  One way to minimize the effect of these short paths is to make the
openings to the leads very small so that the probability of being trapped
for a long time increases. Presumably the CE will apply to any completely
chaotic billiard in the limit that the openings are very small (the number
of modes in each lead should remain constant). However, such a structure is
difficult to treat numerically, except in the very small $N$ limit, which,
in fact, we discuss in section \ref{numerics with d.p.} below.

\subsection{Absence of direct processes}

\begin{figure}[tb]
   \begin{center}
   \leavevmode
   \epsfxsize = 12.5cm
   \epsfbox{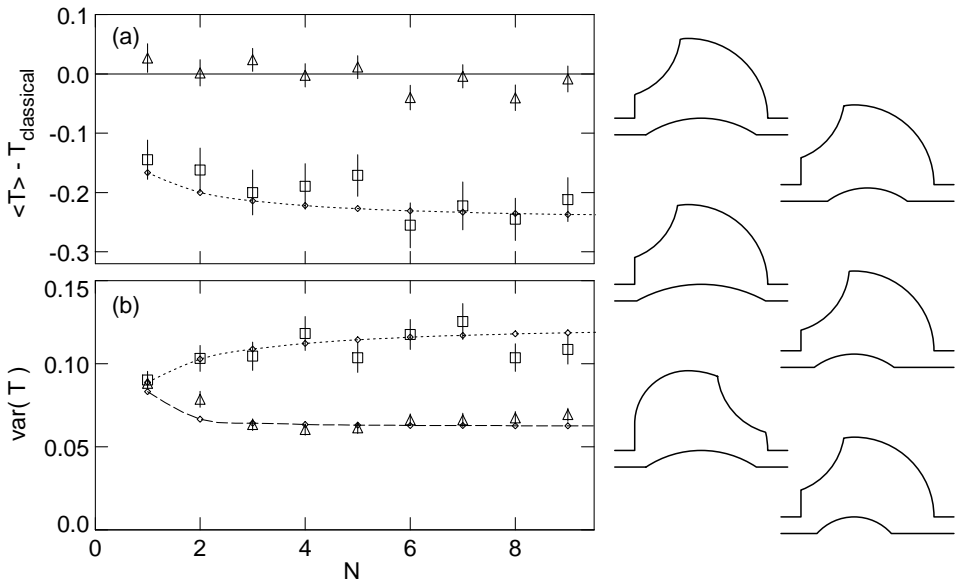}
   \end{center}
   \vspace*{-0.5cm}\caption{
   The magnitude of (a) the quantum correction to the classical conductance
   and (b) the conductance fluctuations as a function of the number of modes
   in the lead $N$. The numerical results for $B \smeq 0$ (squares with
   statistical error bars) agree with the prediction of the COE (points,
   dotted line), while those for $B \smneq 0$ (triangles) agree with the CUE
   (points, dashed line). The six cavities shown on the right were used; the
   average of the numerical results is plotted. Note that each cavity has
   stoppers to block both the direct and whispering gallery trajectories. 
   For non-zero field, $BA/\phi_0  \smeq  2,$ $4$ where $A$ is the
   area of the cavity.
   (After Ref. \protect\cite{BarMelloPRL94}.)}
   \label{num_absence_T}
\end{figure}

In order to compare with the predictions of the circular ensembles, we wish,
therefore, to study structures in which the most obvious direct paths are
absent. To this end, we have introduced ``stoppers'' into the stadium to
block both the direct and the whispering gallery trajectories: examples are
shown in Fig. \ref{num_absence_T}. In order to study the statistical
properties, the conductance as a function of energy is calculated. Because
the energy variation is on the scale of $\hbar \gamma_{\rm esc}$ (the escape
rate from the cavity) \cite{Blu88,JalBarSto90}, it is much more rapid than
the spacing between the modes in the leads ($\hbar v_F/W$). Thus many
independent samplings of the conductance at a fixed number of modes may be
obtained. In addition, we vary slightly the position of the stoppers so as
to change the interference effects and collect better statistics. The
numerical results in Fig. \ref{num_absence_T} used 50 energies for each $N$
(all chosen away from the threshold for the modes) and the 6 different
stopper configurations shown; the classical transmission probabilities for
these cavities ranged from $0.46$ to $0.51$ with a mean of $0.49$.  In
addition, for non-zero magnetic field two values were used.  We see that the
agreement with the CE is now very good for both the mean and the variance,
both for $B \smeq 0$ and for non-zero $B$.  This supports our view that the
deviations in the simple structure Fig. \ref{num_simple_T} are caused by
short paths.

\begin{figure}[bt]
   \begin{center}
   \leavevmode
   \epsfxsize = 0.95\textwidth
   \epsfbox{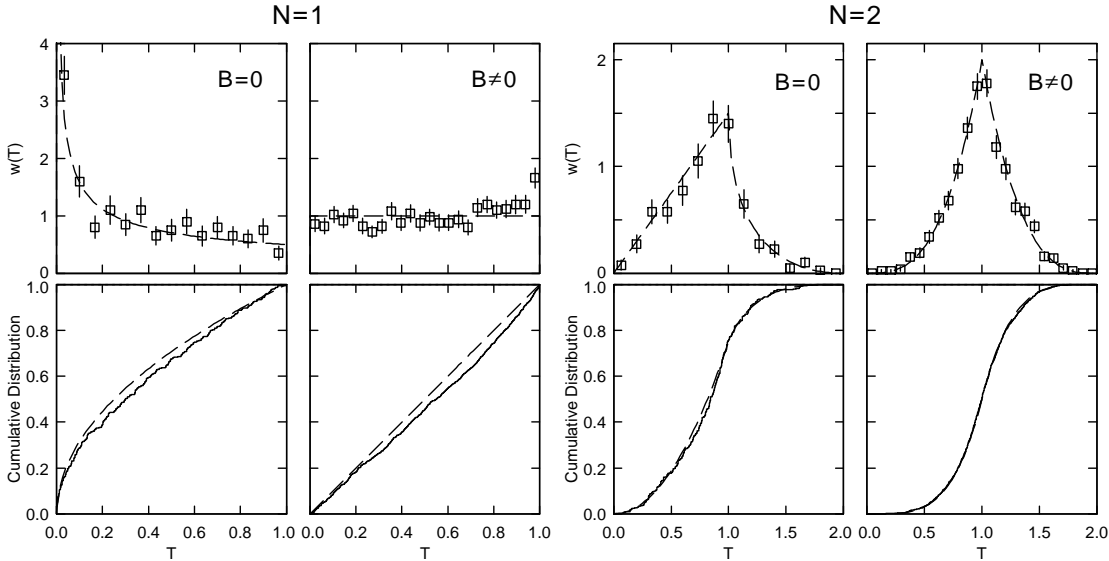}
   \end{center}
   \vspace*{-0.5cm}\caption{
   The distribution of the transmission intensity at fixed $N \smeq 1$ or
   $2$ in both the absence and presence of a magnetic field, compared to
   the analytic COE and CUE results. The panels in the first row
   are histograms; those in the second row are cumulative distributions.
   Note both the strikingly non-Gaussian distributions and the good 
   agreement between the numerical results and the CE in all cases.
   The cavities and energy sampling points used are the same as those in 
   Fig. \protect\ref{num_absence_T}; for $B \smneq 0$, $BA/\phi_0  \smeq  2$,
   $3$, $4$, and $5$ were used.
   (After Ref. \protect\cite{BarMelloPRL94}.)}
   \label{num_absence_w}
\end{figure}

While the agreement of the mean and variance of the conductance with the CE
gives a strong indication of the validity of the information-theoretic model
for real cavities, a much more dramatic prediction of the model is the
strongly non-Gaussian distribution of $T$ for a small number of modes.  The
analytic results for the CE were given in section \ref{w(T)}. These are
compared to the numerical results for $N \smeq 1$, $2$ in Fig.
\ref{num_absence_w}, using the same data as in Fig. \ref{num_absence_T}.
Note that the data is consistent with a square-root singularity in the case
$N \smeq 1$ $B \smeq 0$ and with cusps in the two $N \smeq 2$ cases.  Thus
we see that even for this much more stringent test, the agreement between
the behavior of real cavities and the CE is excellent.

\subsection{Presence of direct processes}
\label{numerics with d.p.}

We now want to look at more general structures than those used in the last
section for comparing to the circular ensemble results. In particular, we
will remove the stoppers that blocked short paths, and compare the numerical
results with the predictions of Poisson's kernel, following Ref.
\cite{BarMelloEPL96}. We will not, however, consider the most general
structure: Poisson's kernel is expected to hold in situations where there
are {\it two} widely separated time scales, a prompt response and an
equilibrated response.  Thus we will study structures where we expect this
to be true.  Since we have only obtained explicit results in the case $N
\smeq 1$ (see section \ref{direct processes}), we will study this case
numerically as well.

\begin{figure}[tb]
   \begin{center}
   \leavevmode
   \epsfxsize = 0.95\textwidth
   \epsfbox{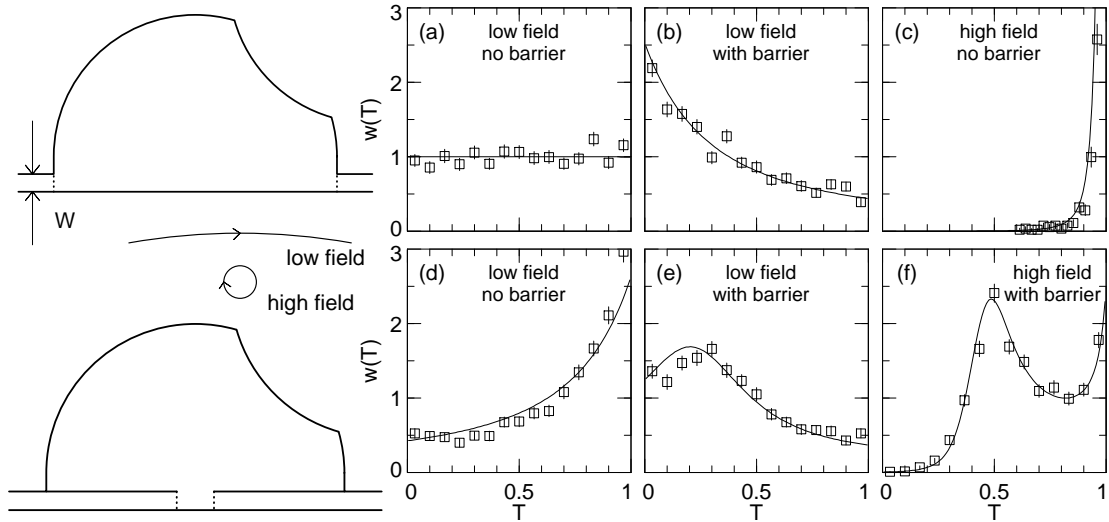}
   \end{center}
   \vspace*{-0.5cm}\caption{
   The distribution of the transmission coefficient for $N \smeq 1$ in
   a simple billiard (top row) and a billiard with leads extended into the
   cavity (bottom row). The magnitude of the magnetic field and the
   presence or absence of a potential barrier at the entrance to the leads
   (marked by dotted lines in the sketches of the structures) are noted in
   each panel. Cyclotron orbits for both fields, drawn to scale, are shown
   on the left. The squares with statistical error bars are the numerical
   results; the lines are the predictions of the information-theoretic
   model, parametrized by an optical $S$-matrix extracted from the
   numerical data. The agreement is good in all cases.
   (From Ref. \protect\cite{BarMelloEPL96}.)}
   \label{num_presence} 
\end{figure}

We have computed the conductance for several billiards shown in Fig.
\ref{num_presence}. Statistics were collected by (1)~sampling in an
energy window larger than the energy correlation length but smaller
than the interval over which the prompt response changes, and
(2)~using several slightly different structures. Typically we used
200 energies in $kW/\pi \in [1.6,1.8]$ (where $W$ is the width of the
lead) and 10 structures found by changing the height or angle of the
convex ``stopper''. Note that the stopper here is used to increase
statistics, not block short paths. As in the absence of direct
processes, since we are mostly averaging over energy, we rely on
ergodicity to compare the numerical distributions to the ensemble
averages of the information-theoretic model. The optical $S$-matrix
was extracted directly from the numerical data and used as $\langle S
\rangle$ in Poisson's kernel; in this sense the theoretical curves
shown below are {\it parameter free}.

We first consider the billiard shown in Fig. \ref{num_presence} at
low magnetic field ($BA/ \phi_0  \smeq  2$ where $A$ is the area of the
cavity; by low magnetic field we mean that the cyclotron radius $r_c$
is much larger than the size of the cavity ($r_c  \smeq  55 \; W$ shown to
scale). In this case $w(T)$ is nearly uniform 
[Fig. \ref{num_presence}(a)], and $\langle S \rangle$ is small because
direct trajectories are negligible in this large structure. We thus
obtain good agreement with the invariant-measure prediction of a
constant distribution (\ref{w(g) N=1}).

In order to increase $\langle S \rangle$ we make one of three changes:
(1)~introduce potential barriers at the openings of the leads into the 
cavity (dashed lines in structures of Fig. \ref{num_presence}), 
(2)~increase the magnetic field, or (3)~extend the leads into the cavity.
In the first case, the barriers are chosen so that the bare transmission 
of each barrier is $1/2$ (determined by calculation in an infinite
lead). They cause direct reflection and skew the distribution towards
small $T$ [Fig. \ref{num_presence}(b)]. Since the reflection from the
barrier is immediate while the transmitted particles are trapped for a
long time, one has two very different response times. 
Second, the large magnetic field ($BA/ \phi_0  \smeq  80$) 
corresponds to $r_c$ just larger than the width of
the lead ($r_c  \smeq  1.4 \: W$). The field increases one component of the 
direct transmission-- the one corresponding to skipping orbits along 
the lower edge-- and skews the distribution towards large $T$ 
[Fig. \ref{num_presence}(c)]. 
Third, extending the leads into the cavity increases the direct
transmission in both directions and also skews the distribution
towards large $T$ [Fig. \ref{num_presence}(d)]. We have done this
rather than consider a smaller cavity since in our case the
equilibrated component is trapped for a long time, yielding a clear
separation of scales.

In each of these cases, the numerical histogram is compared to the
information-theoretic model (solid lines) in which the numerically
obtained $\overline{S}$ is inserted. In panels (b)-(d) the curve plotted 
is the analytic expression of Eq. (\ref{w(T) X,Y beta=2 b}) and 
the corresponding one for direct transmission.
Note the excellent agreement with the information-theoretic model.

Since the long-time classical dynamics in each of the three
structures (a), (b) and (d) is chaotic, these results show that a
wide variety of behavior is possible for chaotic scattering, the
invariant-measure description applying only when there is a single
characteristic time scale.

In case (c), the dynamics is not completely chaotic because of the
small cyclotron radius, and so one would not expect the circular
ensemble to apply. In Ref. \cite{BarMelloPRL94} we found that
increasing the magnetic flux through the structure beyond a few flux
quanta spoiled the agreement with the circular ensemble; we now know
that a nonzero $\langle S \rangle$ is generated and that the present
model describes the data very well. The excellent agreement found
here with a flux as high as 80 suggests extending the analysis to the
quantum Hall regime.

By combining several of the modifications used above, different
$\langle S \rangle$ and so different distributions can be produced.
First, by using extended leads with barriers at their ends, one can
cause both prompt transmission and reflection; the result is shown in
Fig. \ref{num_presence}(e). Finally, increasing the magnetic field in
this structure produces a large average transmission and a large
average reflection. The resulting $w(T)$, Fig. \ref{num_presence}(f),
has a surprising two-peak structure: one peak near $T \smeq 1$ caused by
the large direct transmission and another near $T \smeq 1/2$. For cases
(e) and (f), four intervals of 50 energies each were treated
independently (since the four intervals show slightly different
$\langle S \rangle$'s ) and the four sets of data were then
superimposed \cite{superpose intervals}. 
Even in these two unusual cases, the prediction of the
information-theoretic model is in excellent agreement with the
numerical results.

\section{Comparison with experimental data: Dephasing effects}
\label{dephasing}

The random $S$-matrix theory of quantum transport through ballistic chaotic
cavities was seen in the last section to be in good agreement
with numerical simulations for structures in which the assumptions of the
theory are expected to hold: that agreement includes the average conductance,
its variance and probability density. 

The ultimate test of the theory should, however, be comparisons with
experiment. A na\"{\ii}ve comparison indicates poor agreement.
For instance, for $N_{1} \smeq N_{2} \smeq N \smeq 2$, both the
theoretical weak-localization correction (WLC) and variance are
larger than the experimental results \cite{Marcus95}. The theoretical
WLC (including spin factor) is $-2/5 \smeq -0.40$ from Eq. (\ref{<g> beta=1 1}),
while the experimental one is $\sim -0.31$. The theoretical $\rmvar(g)$ is
$4/15 \approx 0.27$ for $\beta  \smeq 2$ [Eq. (\ref{var(g) beta=2})] and
$72/175 \approx 0.41$ for $\beta  \smeq 1$ [Eq. (\ref{var(g) beta=1})], 
giving the ratio 
$\left[ \rmvar(g)\right] _{0}^{(1)}/\left[ \rmvar(g)\right] _{0}^{(2)} \smeq 
54/35 \approx 1.54$. The experimental results for the variance, on the
other hand, are $\sim 0.015$ for $B\neq 0$ and $\sim 0.034$ for $B \smeq 0$, 
giving the ratio $2.27$. In addition, the measured probability density 
\cite{Marcus95} is close to a Gaussian distribution, which differs from the
prediction of Eqs. (\ref{w(g) N=2 beta=1}), (\ref{w(g) N=2 beta=2}). More
recent experimental data for $N \smeq 1$ \cite{Marcus98} show a distribution
which is clearly asymmetric for $B \smeq 0$, favoring low conductances as
required by weak localization, but the asymmetry is by no means as
pronounced as that of Eq. (\ref{w(g) N=1}) which has a square-root
singularity at the origin.

To reconcile these discrepancies,
we must realize that inherent in the discussion of the previous sections
is the assumption that one can neglect processes which destroy the coherence
of the wave function inside the sample and neglect energy smearing caused by
non-zero temperature. Even if the phase-breaking
length $l_{\phi }$ is larger than the geometrical size of the cavity, for
sufficiently narrow leads the electron may spend enough time inside the
sample to feel the effect of phase-breaking mechanisms. The discussion of
these effects and their relevance for the description of the
experimental data is the subject of the present section.

We simulate the presence of phase-breaking events through a model invented
by M. B\"{u}ttiker \cite{buettiker86}, where, in addition to the physical
leads 1,2 attached to reservoirs at chemical potentials $\mu _{1}$, $\mu
_{2} $, a ``fake lead'' 3 connects the cavity to a phase-randomizing
reservoir at chemical potential $\mu _{3}$. The idea is that any
particle which exits the cavity through lead $3$ is replaced by a particle
from the reservoir; since the replacement particle comes from a
reservoir, it is incoherent with respect to the exiting particle, hence the
phase-breaking. Requiring the current in lead $3$
to vanish determines $\mu _{3}$; the two-terminal dimensionless conductance
is then found to be 
\begin{equation}
g\equiv G/(e^{2}/h)=2\left[ T_{21}+\frac{T_{23}T_{31}}{T_{32}+T_{31}}\right]
\;\;.  \label{g with fake lead}
\end{equation}
In this equation, the factor of $2$ accounts for spin explicitly and
$T_{ij}$ is the transmission coefficient for ``spinless electrons''
from lead $j$ to lead $i$. This expression for $g$ has a natural
interpretation: the first term is the coherent transmission and the
second term is the sequential transmission from $1$ to $2$ via $3$.
In terms of the $S$ matrix, $T_{ij}$ is
\begin{equation}
T_{ij}=\sum_{a_{i}=1}^{N_{i}}\sum_{b_{j}=1}^{N_{j}}\left|
S_{ij}^{a_{i}b_{j}}\right| ^{2},  \label{Tij}
\end{equation}
where $N_{i}$ is the number of channels in lead $i$; $S_{ij}^{a_{i}b_{j}}$
is the matrix element in the position $a_{i}$, $b_{j}$ of the $ij$ block of
the $S$ matrix, rows and columns in this block being labeled from $1$ to $%
N_{i}$ and from $1$ to $N_{j}$, respectively. The total number of channels
will be designated by 
\begin{equation}
N_{T}=\sum_{i}N_{i}. 
\end{equation}
Occasionally, we shall use the notation $N_{\phi }$ to designate the number
of {\it phase-breaking channels} $N_{3}$ in the fake lead 3. These $N_{\phi
} $ channels are physically related to the phase-breaking scattering rate $%
\gamma _{\phi }$ via the relation $N_{\phi }/(N_{1}+N_{2})\approx \gamma
_{\phi }/\gamma _{\rm esc}$, where $\gamma _{\rm esc}$ is the escape 
rate from the cavity. This ``fake lead'' model has been used for various
studies of the effect of phase-breaking in mesoscopic systems
\cite{fakelead}.

We now make the assumption that the {\it total} $N_{T}\times N_{T}$
scattering matrix $S$ obeys the distribution (\ref{dP=dmu}) given
by the invariant measure \cite{BarMello95,piet-carlo95}. 
Through this assumption, the effect of the third lead is felt somehow
uniformly in space rather than at any given point. This even-handed
statistical treatment makes the ``fake lead'' approach more physically
reasonable. 
A further generalization to the case of non-ideal coupling of the fake 
lead to the cavity is studied in Ref. (\cite{piet-carlo97}) but will not
be presented here. In the following, we confine the analytical
discussion to the large $N_{\phi}$ limit, then present the results of
numerical random-matrix theory calculations, and finally compare again
to experiment.

\subsection{Large $N_\phi$}
\label{large Nphi}

We find below the average and variance of the conductance when 
$N_{\phi } \gg 1$ \cite{BarMello95}. As we shall see, we need, for
that purpose, the average and the covariance of the transmission
coefficients $T_{ij}$ introduced above.

From Eqs. (\ref{<|Saalpha|^2>}) and (\ref{<SabSab star>}),
the average of $T_{ij}$ ($i\neq j$) is given by 
(as usual, $\beta  \smeq 1,2$)
\begin{equation}
\left\langle T_{ij}\right\rangle _{0}^{(\beta )}=\frac{N_{i}N_{j}}{%
N_{T}+\delta _{\beta 1}},  \label{av(Tij)}
\end{equation}
so that
\begin{eqnarray}
\left\langle T_{21}\right\rangle _{0}^{(\beta )}=\frac{N_{2}N_{1}}{%
N_{T}+\delta _{\beta 1}}\approx \frac{N_{2}N_{1}}{N_{3}}+O\left( \frac{1}{%
N_{3}^{2}}\right)  \label{av(T21)}
\\[0.1in]
\left\langle T_{31}\right\rangle _{0}^{(\beta )}=\frac{N_{3}N_{1}}{%
N_{T}+\delta _{\beta 1}}\approx N_{1}\left[ 1-\frac{N_{1}+N_{2}+\delta
_{\beta 1}}{N_{3}}+O\left( \frac{1}{N_{3}^{2}}\right) \right]
\label{av(T31)}
\\[0.1in]
\left\langle T_{32}\right\rangle _{0}^{(\beta )}=\frac{N_{3}N_{2}}{%
N_{T}+\delta _{\beta 1}}\approx N_{2}\left[ 1-\frac{N_{1}+N_{2}+\delta
_{\beta 1}}{N_{3}}+O\left( \frac{1}{N_{3}^{2}}\right) \right] ,
\label{av(T32)}
\end{eqnarray}
where the $\approx $ sign refers to the situation $N_{3}\gg 1$ while 
$N_{1}, N_{2} \smeq O(1)$.
Turning to the covariance, using 
Eqs. (\ref{<|S12|^2|S34|^2> beta=2})-(\ref{<|S12|^4> beta=2}) 
and (\ref{av(Tij)}), one finds for $\beta  \smeq 2$
\begin{equation}
\left\langle \delta T_{ij}\delta T_{kl}\right\rangle _{0}^{(2)}=
\frac{N_i N_j}{N_{T}^{2}\left( N_{T}^{2}-1\right) } 
\left[ N_{T}^{2}\delta _{ik}\delta_{jl}-N_{k}N_{T}\delta _{jl}
-N_{l}N_{T}\delta_{ik}+N_{k}N_{l}\right]  .
\label{cov beta=2}
\end{equation}
Likewise, using 
Eqs. (\ref{<|S12|^2|S34|^2> beta=1})-(\ref{<|S12|^4> beta=1}) 
for $\beta  \smeq 1$ yields
\begin{eqnarray}
\fl
\left\langle \delta T_{ij}\delta T_{kl}\right\rangle _{0}^{(1)}=\frac{1}{%
N_{T}\left( N_{T}+1\right) ^{2}\left( N_{T}+3\right) } 
\nonumber \\[0.05in]
\times \{N_{i}N_{j}\left( N_{T}+1\right) \left( N_{T}+2\right) \left( \delta
_{ik}\delta _{jl}+\delta _{il}\delta _{jk}\right) +2N_{i}N_{k} 
\delta _{ij}\delta _{kl}
\nonumber \\[0.05in]
+2N_{i}N_{k}N_{l}\delta _{ij}+2N_{i}N_{j}N_{k}\delta _{kl}+2N_{i}N_{T}\left(
N_{T}+1\right) \delta _{ijkl}+2N_{i}N_{j}N_{k}N_{l} 
\nonumber \\[0.05in]
-\left( N_{T}+1\right) [2N_{i}N_{l}\delta _{ijk}+2N_{i}N_{k}\delta
_{ijl}+2N_{i}N_{j}(\delta _{ikl}+\delta _{jkl}) 
\nonumber \\[0.05in]
+N_{i}N_{j}N_{l}\left( \delta _{ik}+\delta _{jk}\right)
+N_{i}N_{j}N_{k}\left( \delta _{il}+\delta _{jl}\right) ]\}.
\label{cov beta=1}
\end{eqnarray}
Here, a $\delta $ with two or more indices vanishes unless all its indices
coincide, in which case its value is 1.
These expressions are valid for arbitrary $N_{i}$, $N_{j},$ and also for $%
i \smeq j $ if $T_{ii}$ is interpreted as the reflection coefficient $R_{ii}$ 
from lead $i$ back to itself.

The conductance of Eq. ($\ref{g with fake lead}$) can be expressed as 
\begin{eqnarray}
g & =2\left[ \overline{T}_{21}+\delta T_{21}+\frac{\left( \overline{T}%
_{23}+\delta T_{23}\right) \left( \overline{T}_{31}+\delta T_{31}\right) }{%
\left( \overline{T}_{32}+\delta T_{32}+\overline{T}_{31}+\delta
T_{31}\right) }\right] 
\nonumber \\[0.1in]
& =2\left[ \overline{T}_{21}+\delta T_{21}+\frac{\overline{T}_{23}\overline{T}%
_{31}}{\overline{T}_{32}+\overline{T}_{31}}\frac{\left( 1+ \delta T_{23}/
\overline{T}_{23}\right) \left( 1+ \delta T_{31}/\overline{T}_{31}
\right) }{1+(\delta T_{32}+\delta T_{31})/(\overline{T}_{32}+\overline{T}%
_{31})}\right] ,  \label{g with fake lead 1}
\end{eqnarray}
where we have written 
\begin{equation}
T_{ij}=\overline{T}_{ij}+\delta T_{ij}.  \label{Tij 1}
\end{equation}
For convenience, we use interchangeably a bar or the bracket 
$\left\langle \cdots \right\rangle _{0}^{(\beta )}$ to indicate an average 
over the invariant measure.

From Eqs. (\ref{cov beta=2}) and (\ref{cov beta=1}) we find, for the
variance of $T_{ij}$ ($i\neq j$) 
\begin{equation}
\left\langle \left( \delta T_{ij}\right) ^{2}\right\rangle _{0}^{(2)}=\frac{%
N_{i}N_{j}\left( N_{T}-N_{i}\right) \left( N_{T}-N_{j}\right) }{%
N_{T}^{2}\left( N_{T}^{2}-1\right) }  \label{var(Tij) beta=2}
\end{equation}
and 
\begin{equation}
\left\langle \left( \delta T_{ij}\right) ^{2}\right\rangle _{0}^{(1)}=\frac{%
N_{i}N_{j}\left[ \left( N_{T}+1-N_{i}\right) \left( N_{T}+1-N_{j}\right)
-\left( N_{T}+1\right) +N_{i}N_{j}\right] }{N_{T}\left( N_{T}+1\right)
^{2}\left( N_{T}+3\right) }.  \label{var(Tij) beta=1}
\end{equation}
For two leads only, i.e. for $N_{\phi } \smeq 0$, these last two expressions
reduce to Eqs. (\ref{var(g) beta=2}) and (\ref{var(g) beta=1}),
respectively. On the other hand, for $N_{\phi }\gg 1$ and $i \smeq 1,2$
\begin{equation}
\frac{\left[ \left\langle \left( \delta T_{i3}\right) ^{2}\right\rangle
_{0}^{(\beta )}\right] ^{1/2}}{\overline{T}_{i3}}\sim O\left( \frac{1}{%
N_{\phi }}\right) .  \label{delta(T13)/<T13>}
\end{equation}
Thus, 
$\delta T_{23}/\overline{T}_{23}$ and $\delta T_{13}/\overline{T}_{13}$ 
are small quantities in Eq. (\ref{g with fake lead 1}), 
and we can make the expansion 
\begin{eqnarray}
\fl
g=2\left( \overline{T}_{21}+\delta T_{21}\right) +2\frac{\overline{T}_{23}%
\overline{T}_{31}}{\overline{T}_{32}+\overline{T}_{31}}
\left[1+\frac{\delta T_{23}%
}{\overline{T}_{23}}+\frac{\delta T_{31}}{\overline{T}_{31}} 
-\frac{\delta T_{32}+\delta T_{31}}{\overline{T}_{32}+\overline{T}_{31}}
+\frac{\delta T_{23}\delta T_{31}}{\overline{T}_{23}\overline{T}_{31}}
\right.
\nonumber \\[0.1in]
\lo+\left.\frac{\left( \delta T_{23}+\delta T_{31}\right) ^{2}}
      {\left( \overline{T}_{32}+\overline{T}_{31}\right) ^{2}} 
-\frac{\delta T_{23}\left( \delta T_{32}+\delta T_{31}\right) }
      {\overline{T}_{23}\left( \overline{T}_{32}+\overline{T}_{31}\right) }
-\frac{\delta T_{31}\left( \delta T_{32}+\delta T_{31}\right) }
      {\overline{T}_{31}\left( \overline{T}_{32}+\overline{T}_{31}\right) }
+\cdots \right] .
\label{g with fake lead 2}
\end{eqnarray}
The average of this conductance for $N_{\phi} \gg 1$ is obtained
by using Eq. (\ref{delta(T13)/<T13>}):
\begin{equation}
\left\langle g\right\rangle _{0}^{(\beta )}=2\left\{ \overline{T}_{21}
+\frac{\overline{T}_{23}\overline{T}_{31}}{\overline{T}_{32}
+\overline{T}_{31}}\left[ 1+O\left( \frac{1}{N_{\phi }^{2}}
\right) \right] \right\} .
\label{<g> with fake lead 2}
\end{equation}
Substituting Eqs. (\ref{av(T21)})-(\ref{av(T32)}), we have 
\begin{equation}
\left\langle g\right\rangle _{0}^{(\beta )}=
2\frac{N_{2}N_{1}}{N_{2}+N_{1}}\left[ 1-\frac{\delta _{\beta 1}}{N_{\phi }}%
\right] +O\left( \frac{1}{N_{\phi }^{2}}\right)
\label{<g> with fake lead 3}
\end{equation}
which gives the weak-localization correction 
\begin{equation}
\delta \left\langle g\right\rangle \equiv \left\langle g\right\rangle
_{0}^{(1)}-\left\langle g\right\rangle _{0}^{(2)}=-2\frac{N_{2}N_{1}}{%
N_{2}+N_{1}}\frac{1}{N_{\phi }}+O\left( \frac{1}{N_{\phi }^{2}}\right) .
\label{WLC}
\end{equation}

Turning to the variance, we subtract the average conductance from
Eq. (\ref{g with fake lead 2}) and obtain to lowest order in 
$1/N_{\phi }$
\begin{equation}
\frac{1}{2}\delta g \equiv 
\frac{1}{2}\left[ g-\left\langle g\right\rangle _{0}^{(\beta )}\right]
\approx \delta T_{21}+\frac{\overline{T}_{23}\overline{T}_{31}}
{\overline{T}_{32}+\overline{T}_{31}}
\left[
\frac{\delta T_{23}}{\overline{T}_{23}} 
+\frac{\delta T_{31}}{\overline{T}_{31}}
-\frac{\delta T_{32}+\delta T_{31}}{\overline{T}_{32}+\overline{T}_{31}}
\right].
\label{g-<g> with fake lead}
\end{equation}
and obtain, for $\beta  \smeq 1,2$, to lowest order in $1/N_{\phi }$ 
\begin{equation}
\frac{1}{2}\delta g \approx \delta T_{21}
+\frac{N_{1}}{N_{2}+N_{1}}\delta T_{23}+\left( \frac{N_{2}}{N_{2}+N_{1}}%
\right) ^{2}\delta T_{31}-\frac{N_{2}N_{1}}{\left( N_{2}+N_{1}\right) ^{2}}%
\delta T_{32} .  \label{g-<g> 1}
\end{equation}
Squaring this last expression and averaging, we obtain the variance of
the conductance in terms of the $\langle \delta T_{ij} \delta T_{kl}
\rangle_0$.
We must now substitute the variances and covariances given in 
Eqs. (\ref{cov beta=2}) and (\ref{cov beta=1}) to obtain our final
result:
\begin{equation}
\left[ \rmvar(g)\right] _{0}^{(\beta )}=\left[ \frac{2N_{2}N_{1}}{\left(
N_{2}+N_{1}\right) N_{\phi }}\right] ^{2}\frac{2}{\beta }
\left[ 1+(2-\beta )\frac{N_{2}^{3}+N_{1}^{3}}{N_{2}N_{1}\left(
N_{2}+N_{1}\right) ^{2}}\right] +\cdots .  
\label{var(g) 1}
\end{equation}
The ratio of variances for $\beta  \smeq 1$ and $\beta  \smeq 2$ is 
\begin{equation}
\frac{\left[ \rmvar(g)\right] _{0}^{(1)}}{\left[ \rmvar(g)\right] _{0}^{(2)}}
=2\left[ 1+\frac{N_{2}^{3}+N_{1}^{3}}{N_{2}N_{1}\left( N_{2}+N_{1}\right)
^{2}}\right] +\cdot \cdot \cdot .  \label{ratio of var}
\end{equation}
We observe from this last equation, first, that this ratio is independent of
$N_{\phi}$ to leading order, and, second, that the ratio of variances is larger
than 2, and as high as 3 for $N_{2} \smeq N_{1} \smeq N \smeq 1$. For comparison, 
in the absence of
phase-breaking and for $N_{2} \smeq N_{1}$, that ratio lies between 1 and 2.

\subsection{Arbitrary $N_{\phi}$}

\begin{figure}[tb]
   \begin{center}
   \leavevmode
   \epsfxsize = 7.0cm
   \epsfbox{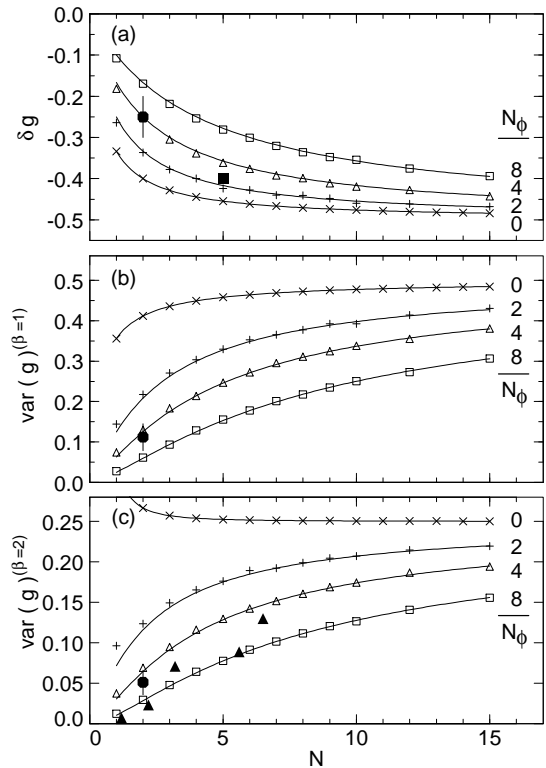}
   \end{center}
   \vspace*{-0.5cm}\caption{
   Magnitude of quantum transport effects as a function of the number of
   channels in the leads, 
   $N_1 \smeq N_2 \smeq N$, for $N_{\phi} \smeq 0$, $2$, $4$,
   and $8$. (a) The weak-localization correction. (b) The variance for
   the orthogonal case ($B\smeq0$). (c) The variance for the unitary case
   (nonzero $B$). Open symbols are numerical results ($20,000$
   matrices used, statistical error is the symbol size). Solid lines
   are interpolation formulae. Solid symbols are experimental results
   of Refs.
   \protect\cite{Chang} (squares), 
   \protect\cite{Westervelt} (triangles), and
   \protect\cite{Marcus95} (circles) corrected for thermal averaging.
   The introduction of phase-breaking decreases the ``universality'' of
   the results but leads to good agreement with experiment.
   (From Ref. \protect\cite{BarMello95}.)
   }
   \label{num_phase_g} 
\end{figure}

In order to evaluate effects of phase-breaking when $N_{\phi}$ is not
large, we numerically evaluate the random-matrix theory, generating
random $N_T \times N_T$ unitary or orthogonal matrices and computing
$g$ from Eq. (\ref{g with fake lead}). Fig. \ref{num_phase_g} shows
the weak-localization correction and the variance of the conductance
as the number of modes in the leads is varied for several fixed
$N_{\phi}$. This result is relevant to experiments at fixed
temperature in which the size of the opening to the cavity is
varied.  Though $\delta g$ and ${\rm var} g$ are nearly independent
of $N$ in the perfectly coherent limit-- the ``universality'' discussed
above-- phase-breaking channels cause variation. Thus
the universality can be seen only if $N_{\phi} \ll N$; otherwise, the
behavior is approximately linear.

The results of Sec. \ref{no direct processes}, which correspond to $N_{\phi
}\smeq0$, and those of the last section, obtained for $N_{\phi }\gg 1$,
suggest an approximate interpolation formula for arbitrary values of 
$N_{\phi }$. For the case $N_{2}\smeq N_{1}\smeq N$, Ref. \cite{BarMello95}
proposes 
\begin{equation}
\delta \left\langle g\right\rangle \approx -\frac{N}{2N+N_{\phi }}
\label{interpolation WLC}
\end{equation}
for the WLC and 
\begin{equation}
\left[ \rmvar(g)\right] _{0}^{(\beta )}\approx \left\{ \left[ \left[
\rmvar(g)\right] _{0,N_{\phi }=0}^{(\beta )}\right] ^{-1/2}+\left[ \left[
\rmvar(g)\right] _{0,N_{\phi }\gg 1}^{(\beta )}\right] ^{-1/2}
\right\}^{-2}
\label{interpolation var(g)}
\end{equation}
for the variance.
These interpolation formulae are compared with the numerical simulations
in Fig. \ref{num_phase_g}: the agreement is good, the only significant
deviation being for $N\smeq 1$ and small $N_{\phi }$. 

\begin{figure}[tb]
   \begin{center}
   \leavevmode
   \epsfxsize = 9.0cm
   \epsfbox{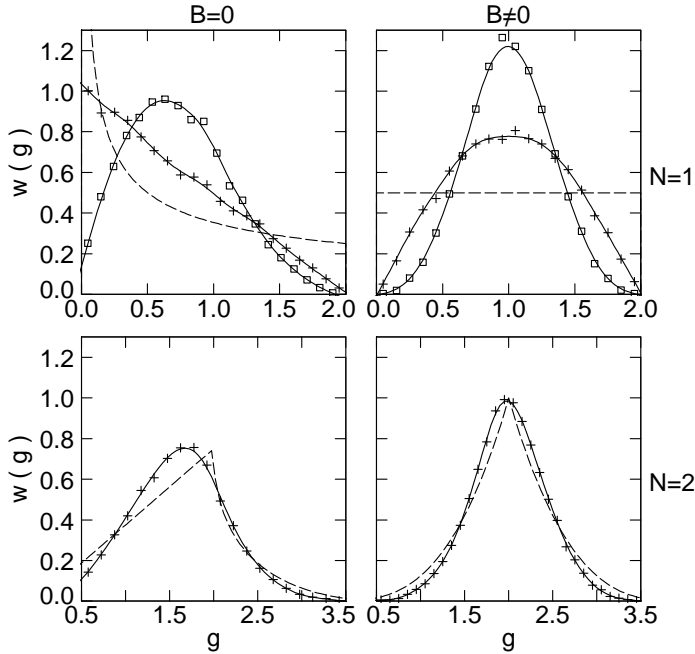}
   \end{center}
   \vspace*{-0.5cm}\caption{
   Probability density of the conductance in the orthogonal (first
   column) and unitary (second column) cases for $N\smeq 1$ (first row) and
   $N\smeq 2$
   (second row).  Increasing the phase-breaking from zero (dashed lines,
   analytic) to $N_{\phi} \smeq 1$ (plus symbols, numerical) to 
   $N_{\phi}\smeq 2$
   (squares, numerical) smooths the distribution.
   (From Ref. \protect\cite{BarMello95}.)
   }
   \label{num_phase_w} 
\end{figure}

In addition to the mean and variance, the probability density of the
conductance, $w (g)$, can be evaluated. Fig. \ref{num_phase_w} shows
the distribution in the weak phase-breaking regime with a small
number of modes in the lead. From the variance and mean, we know that
the phase-breaking will narrow the distribution and make it more
symmetric because the weak-localization correction goes to zero. 
Not surprisingly, the phase-breaking in addition smooths
the distribution, and the extreme non-Gaussian structure in $w(g)$ is
washed out.

\subsection{Experiment}

Now that we have a tool for evaluating the effects of phase-breaking, we
return to a comparison of the information-theoretic model with
experiments. First, for the mean and variance of the conductance, in Fig.
\ref{num_phase_g} are also shown as black dots the results of three
experiments from Refs. \cite{Chang,Westervelt,Marcus95}.  Since the
theoretical results take into account through $ N_{\phi }$ the effect of
finite temperature insofar as dephasing is concerned, without including
thermal smearing, the experimental variance has been corrected to
compensate for this latter effect \cite{correct exp var}.  In the case of
Ref. \cite{Marcus95} measurements of all three quantum transport
quantities were made, and so this data can be used to test the consistency
of the theory.  Notice that a value of $N_{\phi }\smeq 4$-$8$ allows the
simultaneous fit of the WLC and the variance for both $\beta  \smeq 1$ and
$2$. This indicates that the theory and experiment are indeed in good
agreement.

\begin{figure}[tb]
   \begin{center}
   \leavevmode
   \epsfxsize=10.5cm
   \epsffile[50 240 550 630]{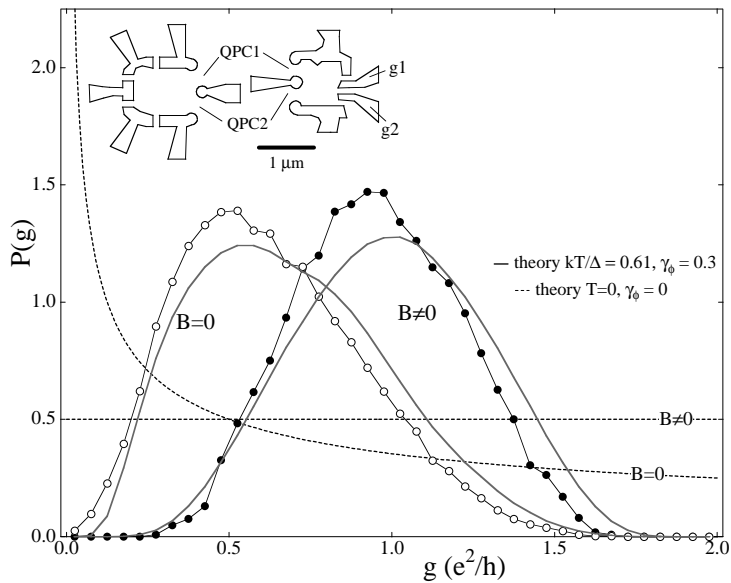}
   \end{center}
   \vspace*{-0.5cm}\caption{
   Experimental conductance distributions for both $B \smeq 0$ (open circles)
   and $40$ mT (filled circles) for a $0.5$ $\mu{\rm m}^2$ device at $100$ mK 
   with $N \smeq 1$. These are compared to theoretical curves for both 
   zero temperature (dashed lines) and non-zero temperature (solid lines). 
   Though the effect of non-zero temperature is substantial, the measured 
   distribution at $B \smeq 0$ is clearly not Gaussian. 
   Inset: pattern of gates defining the dot.
   (After Ref. \protect\cite{Marcus98}.)
   }
   \label{phase_expt} 
\end{figure}

Second, the full distribution of the conductance has been measured in
the $N \smeq 1$ case 
and compared to the random-matrix theory including phase-breaking
\cite{Marcus98}. The results are shown in Fig. \ref{phase_expt}.
In the $B \smeq 0$ case,
the experimental distribution is clearly not Gaussian and is skewed towards
small values of $g$, as expected. 
On the other hand, the distribution is not nearly as dramatic as the
completely coherent theory suggests.
An important recent development in relation with the fictitious lead model
described above has been carried out in Ref. \cite{piet-carlo97}, with the
motivation of describing a spatially more uniform distribution of
phase-breaking events: the fictitious lead is considered to support an
infinite number of modes, each with vanishing transmission, allowing a
continuous value for the dephasing rate. The experimental results 
are analyzed in Ref. \cite{Marcus98} with this improved model, and the
result is also shown in Fig. \ref{phase_expt}. It is seen that the shape 
of the conductance distribution is reproduced well with a dephasing rate 
obtained independently from the WLC. This comparison is strong support for 
the validity of the theory. The numerical value of the variance 
at various temperatures shows a
discrepancy, though; in particular, the observed ratio of variances for 
$\beta  \smeq 1$ and $\beta  \smeq 2$ is considerably larger than that given 
by the model: this is as yet unexplained.

\section{Conclusion}
\label{conclusion}

In this paper we developed a statistical theory aimed at the description
of the quantum-mechanical scattering of a particle by a cavity, whose
geometry is such that the classical dynamics of the system is chaotic. We
studied, as our main application, the electronic transport through
ballistic microstructures, in an independent-particle approximation.

The theory, which was developed in the past within nuclear physics to
describe the scattering of a nucleon by a nucleus, describes the regime in
which there are two distinct time scales, associated with a prompt and an
equilibrated response. The prompt response is described in terms of the
energy average $\overline{S}$-- also known as the optical $S$ matrix-- of the
actual $S$ matrix. Through the notion of ergodicity, $\overline{S}$ is
calculated as the average $\left\langle S\right\rangle $ over an ensemble of
similar systems, represented mathematically by an ensemble of $S$ matrices
belonging to the universality class in question: orthogonal, unitary or
symplectic. In addition, the ensemble satisfies the analyticity
(causality)-ergodicity requirements and the optical $\left\langle
S\right\rangle $ evaluated over it has a specified (matrix) value. The
ensemble discussed in the text is the one that carries minimum information,
or maximum entropy, satisfying these conditions. It is thus meant to describe
those situations in which any other additional information is irrelevant. In
this procedure one constructs the one-energy statistical distribution of $S$
using only the above physical information-- expressible in terms of $S$
itself-- without ever invoking any statistical assumption for the underlying
Hamiltonian which never enters the analysis.

From the resulting $S$-matrix distribution, known as Poisson's kernel,
properties of the quantum conductance have been derived: its average, its
fluctuations, and its full distribution in certain cases, both in the absence
and presence of direct processes. We obtain good agreement with the results
of the numerical solution of the Schr\"{o}dinger equation for cavities in
which the assumptions of the theory hold: either ones in which prompt
response is absent (section \ref{no direct processes}), or ones in which
there are two widely separated time scales (section \ref{direct processes}).
As for the comparison with experimental data, agreement was found once
temperature smearing and dephasing effects were taken into account, at least
within a phenomenological model.

The effect of time-reversal symmetry-- determining the universality class
$\beta $-- has been seen to be of fundamental importance. One other symmetry
was not touched upon in this article, because of lack of space: {\it spatial
reflection symmetries}. While such symmetries are not relevant for
disordered systems-- the traditional subject of mesoscopic physics-- they are
possible in ballistic systems in which one can control the scattering
geometry. How such symmetries affect the interference contribution to
transport was studied in Refs. \cite{BarMello96,GoparJPA96}. For reflection
symmetries, $S$ is block diagonal in a basis of definite parity with respect
to that reflection, with a circular ensemble in each block (if $\left\langle
S\right\rangle \smeq 0$). The key point is that the conductance may couple
the different parity-diagonal blocks of $S$, and thus the resulting quantum
transport properties are a nontrivial generalization of the circular ensemble
results. These effects are discussed in Ref. \cite{BarMello96,GoparJPA96} for
structures presenting a ``left-right'', ``up-down'', ``four-fold'', and
``inversion'' spatial symmetry and compared with the results of numerical
solutions of the Schr\"{o}dinger equation. From an experimental point of
view, it would be nice to have a confirmation of our results, both in
microstructures as well as in microwave cavities, the latter ones being
perhaps simpler to handle.

We should point out that the theory presented here is not applicable when
there are other relevant time scales in the problem. One example is the case
of a disordered quasi-1D system, where the diffusion time across the system
is also an important time scale. In the past, this extra piece of
information was taken into account explicitly by expressing the transfer
matrix for the full system as the product of the transfer matrices for the
individual slices that compose the system
\cite{Dorokhov,MelPeyKumar,mello-stone,AltLeeWebb91,LesHmeso,BeenRMP}. An
example of an intermediate situation is that of a finite number of cavities
connected in series. Whether more than two time scales are physically
relevant for {\it single cavities} as well is not clear at present. It seems
likely that the deviations of the numerical results from the theory for the
simple structures in Section \ref{simple structures} are for this reason
\cite{notPoisKern}. Also, it is conceivable that additional time scales are
at the root of the discrepancies discussed in Ref. \cite{BarMello96} in
relation with the WLC. An extension of the maximum-entropy theory presented
above to include other time scales is not known at present.

We also stress that the theory presented here is applicable to the
statistics of functions of the $S$ matrix at a single value of the energy.
The joint statistical distribution of the $S$ matrix at two or more energies
has escaped, so far, an analysis within the philosophy described above (some
aspects of the two-point problem have been studied assuming an underlying
Hamiltonian described by a Gaussian ensemble as in Refs.
\cite{lewenkopf-weidenmuller,fyodorov-sommers}). An approach coming close to
that philosophy was initiated in Ref. \cite{gopar et al} for the simplest
quantity of a two-point character: the statistical distribution of the time
delay arising in the scattering process-- a quantity involving the energy
derivative of $S$-- motivated by the study of the electrochemical
capacitance of a mesoscopic system. It was found that another piece of
information is needed for the description: the statistical distribution of
the $K$-matrix of resonance widths. Ref. \cite{gopar et al} deals with $1
\times 1$ $S$-matrices; subsequent work derived the distribution of time
delays for $N$ channels \cite{BrouFraBeen}. In a further development, Ref.
\cite{gopar-mello} finds a transformation that relates the $k$-point
distribution of the $n$-dimensional $S$-matrix for the case $\left\langle
S\right\rangle \neq 0$ to that for $\left\langle S\right\rangle  \smeq 0$,
thus relating the problems in the presence and in the absence of direct
processes.

Finally, we remark that the information-theory described in the text makes
use of the standard Boltzmann-Shannon definition of information and
entropy.  Other definitions have been presented in the literature: see, for
instance, Ref. \cite{tsallis}, and the references contained therein. What
would be the physics behind the use of other definitions of entropy and how
our results would be modified is not known at present.

\ack
The authors are grateful for helpful conversations with P. W. Brouwer, 
R. A.  Jalabert, V. A. Gopar, C. M. Marcus, M. Mart\'{\ii}nez, and 
T. H. Seligman.
PAM acknowledges partial support of CONACyT, under contract No. 2645P-E.
Both authors wishes to thank the hospitality of The Aspen Center for
Physics, where part of this paper was developed and discussed.

\appendix
\def\thesection{\Alph{section}}

\section{Appendix: Evaluation of the invariant measure}
\label{app inv measure}

We evaluate $dS$ in the polar representation (\ref{S polar repr}) and then
the arc element $ds^2$ of Eq. (\ref{arc element for S}) keeping $V$ and $W$
independent, as is the case for $\beta  \smeq 2$. We use this same algebraic
development and set $W \smeq V^T$ in the proper place to analyze the 
$\beta \smeq 1$ case. Differentiating $S$ of Eq. (\ref{S polar repr}) we 
obtain 
\begin{equation}
dS=(dV)RW+V(dR)W+VR(dW)
=V\left[ (\delta V)R+dR+R(\delta W)\right]  \label{dS}
\end{equation}
where we have defined the matrices 
\begin{equation}
\delta V=V^{\dagger }dV,\qquad \delta W=(dW)W^{\dagger },  \label{deltaV,W}
\end{equation}
which are {\it antihermitian}, as can be seen by differentiating the
unitarity relations $V^{\dagger }V \smeq I$, $WW^{^{\dagger }} \smeq I$. 
The arc element of Eq. (\ref{arc element for S}) is thus 
\begin{eqnarray}
\fl
ds^2=\rmTr\{\left[ R^T(\delta V)^{\dagger }+dR^T+(\delta W)^{\dagger
}R^T\right] 
\cdot \left[ (\delta V)R+dR+R\delta W\right] \}
\nonumber \\[0.03in]
\lo=
\rmTr[R^T(\delta V)^{\dagger }(\delta V)R+R^T(\delta V)^{\dagger
}dR+R^T(\delta V)^{\dagger }R(\delta W)
\nonumber \\[0.03in]
\lo
+(dR^T)(\delta V)R+(dR^T)dR
+(dR^T)R(\delta W)
\nonumber \\[0.03in]
\lo
+(\delta W)^{\dagger }R^T(\delta V)R+(\delta W)^{\dagger }R^TdR+(\delta
W)^{\dagger }R^TR(\delta W)].  
\label{ds2 1}
\end{eqnarray}
The various differentials occurring in the previous equation can be
expressed as 
\begin{eqnarray}
dR= \frac{1}{2} \left[ 
\begin{array}{cc}
d\tau / \sqrt{\rho } & d\tau / \sqrt{\tau } \\ 
d\tau / \sqrt{\tau } & -d\tau / \sqrt{\rho }
\end{array}
\right] ,  \label{dR}
\\[0.05in]
\delta V=\left[ 
\begin{array}{cc}
v_1^{\dagger }dv_1 & 0 \\ 
0 & v_2^{\dagger }dv_2
\end{array}
\right] =\left[ 
\begin{array}{cc}
\delta v_1 & 0 \\ 
0 & \delta v_2
\end{array}
\right] ,  \label{del V}
\\[0.05in]
\delta W=\left[ 
\begin{array}{cc}
(dv_3)v_3^{\dagger } & 0 \\ 
0 & (dv_4)v_4^{\dagger }
\end{array}
\right] =\left[ 
\begin{array}{cc}
\delta v_3 & 0 \\ 
0 & \delta v_4
\end{array}
\right] ,  \label{del W}
\end{eqnarray}
where we have used the abbreviation
\begin{equation}
\rho =1-\tau .  \label{rho}
\end{equation}

We now calculate the various terms in (\ref{ds2 1}). The first plus ninth
terms give 
\begin{equation}
\rmTr\left[ (\delta V)^{\dagger }(\delta V)+(\delta W)^{\dagger }(\delta
W)\right] 
=\rmTr\sum_{i=1}^4\left( \delta v_i\right) ^{\dagger }\delta v_i.
\label{one plus nine}
\end{equation}
The second plus fourth terms give 
\begin{equation}
\fl \rmTr\left[ R^T(\delta V)^{\dagger }dR+(dR^T)(\delta V)R\right] 
\label{two plus four}
\end{equation}
\[
=\rmTr[\sqrt{\rho }\left( \delta v_1\right) ^{\dagger }d\sqrt{\rho }+\sqrt{\tau
}\left( \delta v_2\right) ^{\dagger }d\sqrt{\tau }
+\sqrt{\tau }\left( \delta v_1\right) ^{\dagger }d\sqrt{\tau }+\sqrt{\rho }%
\left( \delta v_2\right) ^{\dagger }d\sqrt{\rho }+h.c.]=0,
\]
where h.c. stands for Hermitian conjugate. We have used the identity 
\begin{equation}
\rmTr\left[ D(\delta v)^{\dagger }D^{\prime }+D^{\prime }(\delta v)D\right] =0,
\label{D(deltav)D'}
\end{equation}
where $D$ and $D^{\prime }$ denote any two real diagonal matrices and $%
\delta v$ an antihermitian one. 
Using the identity 
\begin{equation}
\rmTr\left[ \left( \delta v_i\right) ^{\dagger }D\left( \delta v_j\right)
D\right] =\rmTr\left[ D\left( \delta v_j\right) ^{\dagger }D\left( \delta
v_i\right) \right] ,  \label{deltavDdeltavD}
\end{equation}
we find for the third plus seventh terms
\begin{eqnarray}
\fl
\rmTr\left[ R^T(\delta V)^{\dagger }R(\delta W)+(\delta W)^{\dagger }R^T(\delta
V)R\right] 
\nonumber \\[0.03in]
\lo
=2\rmTr[\sqrt{\rho }\left( \delta v_1\right) ^{\dagger }\sqrt{\rho }\left(
\delta v_3\right) +\sqrt{\tau }\left( \delta v_2\right) ^{\dagger }\sqrt{%
\tau }\left( \delta v_3\right) 
\nonumber \\[0.03in]
\lo
+\sqrt{\tau }\left( \delta v_1\right) ^{\dagger }\sqrt{\tau }\left( \delta
v_4\right) +\sqrt{\rho }\left( \delta v_2\right) ^{\dagger }\sqrt{\rho }%
\left( \delta v_4\right) ].  
\label{three plus seven}
\end{eqnarray}
The fifth term gives 
\begin{equation}
\rmTr(dR^T)(dR)=\frac 12\sum_a\frac{\left( d\tau _a\right) ^2}{\tau _a\rho _a}.
\label{five}
\end{equation}
Finally, the sixth plus eighth terms give 
\begin{eqnarray}
\fl
\rmTr\left[ (dR^T)R(\delta W)+h.c.\right] 
\nonumber \\
\lo
=\frac 12\rmTr\left[ -(d\tau )\delta v_3+(d\tau )\delta v_3+(d\tau )\delta
v_4-(d\tau )\delta v_4+h.c.\right] =0.  \label{six plus eight}
\end{eqnarray}

Substituting these expressions in (\ref{ds2 1}), we find 
\begin{eqnarray}
\fl
ds^2=2 \rmTr\left\{\frac{1}{2} \sum_{i=1}^4
\left( \delta v_i\right) ^{\dagger }(\delta v_i)+
\frac{\left( d\tau \right) \left( d\tau \right) }{4\tau \rho }
+\sqrt{\rho }\left( \delta v_1\right) ^{\dagger }\sqrt{\rho }\left( \delta
v_3\right) \right.
\nonumber \\
\lo+
\left.
\sqrt{\tau }\left( \delta v_2\right) ^{\dagger }\sqrt{\tau }%
\left( \delta v_3\right) 
+\sqrt{\tau }\left( \delta v_1\right) ^{\dagger }\sqrt{\tau }\left( \delta
v_4\right) +\sqrt{\rho }\left( \delta v_2\right) ^{\dagger }\sqrt{\rho }%
\left( \delta v_4\right) \right\}.  \label{ds2 2}
\end{eqnarray}
The antihermitian matrix $\delta v_i$ can be expressed as 
\begin{equation}
\delta v_i=\delta a_i+i\delta s_i,  \label{deltav a is}
\end{equation}
where $\delta a_i$ is real antisymmetric and $\delta s_i$ is real symmetric.
Substituting in the expression for $ds^2$ and rearranging terms, we find 
\begin{eqnarray}
\fl
ds^2=\sum_a\left\{\sum_{i=1}^4\left( \left( \delta s_i\right) _{aa}\right)
^2+2\rho _a\left[ \left( \delta s_1\right) _{aa}\left( \delta s_3\right)
_{aa}+\left( \delta s_2\right) _{aa}\left( \delta s_4\right) _{aa}\right] 
\right.
\nonumber \\
\lo+
\left. 2\tau _a\left[ \left( \delta s_2\right) _{aa}\left( \delta s_3\right)
_{aa}+\left( \delta s_1\right) _{aa}\left( \delta s_4\right) _{aa}\right] +%
\frac{\left( d\tau _a\right) ^2}{2\tau _a\rho _a}\right\}
\nonumber \\
\lo
+2\sum_{a<b}\left\{\sum_{i=1}^4\left( \left( \delta a_i\right) _{ab}\right)
^2+\sum_{i=1}^4\left( \left( \delta s_i\right) _{ab}\right) ^2
\right.
\label{ds2 3}
\\[0.03in]
+2\sqrt{\rho _a\rho _b}[\left( \delta a_1\right) _{ab}\left( \delta
a_3\right) _{ab}+\left( \delta a_2\right) _{ab}\left( \delta a_4\right) _{ab}
+\left( \delta s_1\right) _{ab}\left( \delta s_3\right) _{ab}+\left( \delta
s_2\right) _{ab}\left( \delta s_4\right) _{ab}]
\nonumber \\[0.03in]
+ \left. 2\sqrt{\tau _a\tau _b}[\left( \delta a_2\right) _{ab}\left( \delta
a_3\right) _{ab}+\left( \delta a_1\right) _{ab}\left( \delta a_4\right) _{ab}
+\left( \delta s_2\right) _{ab}\left( \delta s_3\right) _{ab}+\left( \delta
s_1\right) _{ab}\left( \delta s_4\right) _{ab}]\right\}.  
\nonumber
\end{eqnarray}

\subsection{The case $\beta  \smeq 1$}

In this case, $v_3 \smeq v_1^T$, $v_4 \smeq v_2^T$ so 
$\delta v_3 \smeq delta v_1)^T$, $\delta v_4 \smeq delta v_2)^T$, and hence 
\begin{equation}
\delta a_3=-\delta a_1,\qquad \delta s_3=\delta s_1 ,\qquad
\delta a_4=-\delta a_2,\qquad \delta s_4=\delta s_2.  \label{v4 v2T}
\end{equation}
Substituting in (\ref{ds2 3}), we have 
\begin{eqnarray}
\fl
ds^2=2\sum_a\left\{\left( \left( \delta s_1\right) _{aa}\right) ^2
+\left( \left(
\delta s_2\right) _{aa}\right) ^2+2\tau _a\left( \delta s_1\right)
_{aa}\left( \delta s_2\right) _{aa}
\right.
\nonumber \\
\lo+
\left.
\rho _a\left[ \left( \left( \delta s_1\right) _{aa}\right) ^2+\left( \left(
\delta s_2\right) _{aa}\right) ^2\right] +\frac{\left( d\tau _a\right) ^2}{%
4\tau _a(1-\tau _a)}\right\}
\nonumber \\
\lo+4\sum_{a<b}\left\{\left( \left( \delta a_1\right) _{ab}\right) ^2
+\left( \left(
\delta a_2\right) _{ab}\right) ^2+\left( \left( \delta s_1\right)
_{ab}\right) ^2+\left( \left( \delta s_2\right) _{ab}\right) ^2
\right.
\nonumber \\[0.03in]
\lo+
2\sqrt{\tau _a\tau _b}[\left( \delta s_1\right)_{ab}^{\ }\left( \delta
s_2\right) _{ab}-\left( \delta a_1\right) _{ab}\left( \delta a_2\right) _{ab}]
\nonumber \\[0.03in]
\lo+
\left.
\sqrt{\rho _a\rho _b}\left[ \left( \left( \delta s_1\right) _{ab}\right)
^2+\left( \left( \delta s_2\right) _{ab}\right) ^2-\left( \left( \delta
a_1\right) _{ab}\right) ^2-\left( \left( \delta a_2\right) _{ab}\right)
^2\right] 
\right\} ,  \label{ds2 4}
\end{eqnarray}
or
\begin{eqnarray}
\fl
\frac 12ds^2=\sum_a\left\{(1+\rho _a)\left[ \left( \left( \delta s_1\right)
_{aa}\right) ^2+\left( \left( \delta s_2\right) _{aa}\right) ^2\right]
+2\tau _a\left( \delta s_1\right) _{aa}\left( \delta s_2\right) _{aa}
+\frac{\left( d\tau _a\right) ^2}{4\tau _a\rho _a}\right\}
\nonumber \\[0.03in]
\lo
+2\sum_{a<b}\left\{\left( 1+\sqrt{\rho _a\rho _b}\right) \left[ \left( \left(
\delta s_1\right) _{ab}\right) ^2+\left( \left( \delta s_2\right)
_{ab}\right) ^2\right] 
\right.
\label{ds2 5} \\[0.05in]
+
\left.
\left( 1-\sqrt{\rho _a\rho _b}\right) \left[ \left( \left( \delta
a_1\right) _{ab}\right) ^2+\left( \left( \delta a_2\right) _{ab}\right)
^2\right] 
+2\sqrt{\tau _a\tau _b}[\left( \delta s_1\right) _{ab}\left( \delta
s_2\right) _{ab}-\left( \delta a_1\right) _{ab}\left( \delta a_2\right) _{ab}%
]\right\}.  
\nonumber
\end{eqnarray}

In (\ref{ds2 5}), $\left( \delta s_1\right) _{aa}$, $\left( \delta
s_2\right) _{aa}$  and $\tau _a$ ($a \smeq 1,\cdot \cdot \cdot ,N$) 
contribute $N$
independent variations each, while $\left( \delta s_1\right) _{ab}$, $\left(
\delta s_2\right) _{ab}$, $\left( \delta a_1\right) _{ab}$,$\left( \delta
a_2\right) _{ab}$ ($a<b$) contribute $N(N-1)/2$  each, giving a total of
\begin{equation}
\nu =3N+4\frac{N(N-1)}2=2N^2+N,  \label{ind param}
\end{equation}
which is the correct number of independent parameters for a $2N$-dimensional
unitary symmetric matrix.

The metric tensor appearing in Eq. (\ref{arc element}) has a simple block
structure, consisting of  $1\times 1$ and $2\times 2$ blocks along the
diagonal, as follows. There are  $N$ 2-dimensional blocks with rows and
columns labeled  $\left( \delta s_1\right) _{aa}$, $\left( \delta
s_2\right) _{aa}$:
\begin{equation}
\begin{array}{cc}
 & \begin{array}{cc}
   \left( \delta s_1\right) _{aa} & \left( \delta s_2\right) _{aa}
   \end{array} \\
$ $ \\
\begin{array}{c}
\left( \delta s_1\right) _{aa} \\ 
\left( \delta s_2\right) _{aa}
\end{array} & 
\left[ 
\begin{array}{cc}
1+\rho _a & \tau _a \\ 
\tau _a & 1+\rho _a
\end{array}
\right]   
\end{array}
\label{block delta s1,2 aa}
\end{equation}
and $N$ 1-dimensional blocks $1/4\rho _a\tau _a$ labeled by $d\tau _a$,
giving, altogether, the contribution
\begin{equation}
\prod_{a=1}^N\frac{(1+\rho _a)^2-\tau _a^2}{4\rho _a\tau _a}=\prod
_{a=1}^N\frac 1{\tau _a}  \label{contr 1 to detg}
\end{equation}
to $\det g$. There are $N(N-1)/2$ 2-dimensional blocks with rows and columns
labeled $\left( \delta s_1\right) _{ab}$, $\left( \delta s_2\right) _{ab}$,
\begin{equation}
\begin{array}{cc}
 & \begin{array}{lr}
   \left( \delta s_1\right) _{ab}~~ & ~~\left( \delta s_2\right) _{ab}
   \end{array} \\
$ $ \\
\begin{array}{c}
   \left( \delta s_1\right) _{ab} \\ 
   \left( \delta s_2\right) _{ab}
   \end{array}
&
\left[ 
   \begin{array}{cc}
   1+\sqrt{\rho _a\rho _b} & \sqrt{\tau _a\tau _b} \\ 
   \sqrt{\tau _a\tau _b} & 1+\sqrt{\rho _a\rho _b}
   \end{array}
   \right] \cdot 2,  
\end{array}
\label{block delta s1,2 ab}
\end{equation}
and also $N(N-1)/2$ 2-dimensional blocks with rows and columns labeled 
$\left( \delta a_1\right) _{ab}$, $\left( \delta a_2\right) _{ab}$,
\begin{equation}
\begin{array}{cc}
 & \begin{array}{lr}
   \left( \delta a_1\right) _{ab}~~ & ~~\left( \delta a_2\right) _{ab}
   \end{array} \\
$ $ \\
\begin{array}{c}
   \left( \delta a_1\right) _{ab} \\ 
   \left( \delta a_2\right) _{ab}
   \end{array}
&
\left[ 
   \begin{array}{cc}
   1-\sqrt{\rho _a\rho _b} & -\sqrt{\tau _a\tau _b} \\ 
   -\sqrt{\tau _a\tau _b} & 1-\sqrt{\rho _a\rho _b}
   \end{array}
   \right] \cdot 2,  
\end{array}
\label{block delta a1,2 ab}
\end{equation}
giving, altogether, a contribution
\begin{eqnarray}
\fl
\left\{ \left[ 1+\sqrt{(1-\tau _a)(1-\tau _b)}\right] ^2-\tau _a\tau
_b\right\} \left\{ \left[ 1-\sqrt{(1-\tau _a)(1-\tau _b)}\right] ^2-\tau
_a\tau _b\right\} 
\nonumber \\[0.05in]
\lo= \left[ 2-\tau _a-\tau _b+2\sqrt{(1-\tau _a)(1-\tau _b)}\right] 
\left[ 2-\tau _a-\tau _b-2\sqrt{(1-\tau _a)(1-\tau _b)}\right] 
\nonumber \\[0.05in]
\lo= \left( \tau _a-\tau _b\right) ^2.  \label{contr 2 to detg}
\end{eqnarray}

Multiplying (\ref{contr 1 to detg}) and (\ref{contr 2 to detg}) and taking
the square root, as required by Eq. (\ref{vol element}), we find the result
given in Eq. (\ref{w(tau)}).

\subsection{The case $\beta  \smeq 2$}

We go back to Eq. (\ref{ds2 3}). We can write the single summation in that
equation (except for its last term) as
\begin{eqnarray}
\fl
\sum_a\{\left[ \left( \delta s_1\right) _{aa}+\left( \delta s_3\right)
_{aa}\right] ^2+\left[ \left( \delta s_2\right) _{aa}+\left( \delta
s_4\right) _{aa}\right] ^2
-2\tau _a\left[ \left( \delta s_1\right) _{aa}-\left( \delta s_2\right)
_{aa}\right] \left[ \left( \delta s_3\right) _{aa}-\left( \delta s_4\right)
_{aa}\right] \}
\nonumber \\[0.05in]
=\sum_a\left[ (\delta x_a)^2+(\delta y_a)^2-2\tau _a(\delta z_a)(\delta
x_a-\delta y_a-\delta z_a)\right] ,  \label{single sum}
\end{eqnarray}
where we have defined the combinations
\begin{eqnarray}
\delta x_a=\left( \delta s_1\right) _{aa}+\left( \delta s_3\right) _{aa}
\label{delta x}
\\[0.1in]
\delta y_a=\left( \delta s_2\right) _{aa}+\left( \delta s_4\right) _{aa}
\label{delta y}
\\[0.1in]
\delta z_a=\left( \delta s_1\right) _{aa}-\left( \delta s_2\right) _{aa}.
\label{delta z}
\end{eqnarray}
Notice that in $ds^2$ the $4N$ quantities $\left( \delta s_1\right) _{aa}$, 
$\left( \delta s_2\right) _{aa}$, $\left( \delta s_3\right) _{aa}$, $\left(
\delta s_4\right) _{aa}$  appear only through the $3N$ combinations $\delta
x_a$, $\delta y_a$, $\delta z_a$ : these quantities, together with the $%
d\tau _a$ contribute $4N$ independent variations. The $\left( \delta
s_i\right) _{ab}$ and the  $\left( \delta a_i\right) _{ab}$ for $i \smeq 1,
\cdot \cdot \cdot ,4$ and $a<b$ contribute $4\cdot N(N-1)/2$  each, so 
that we have a total of $4N^2$ variations, which is the correct number of
independent parameters for a $2N$-dimensional unitary matrix.

In terms of independent variations we can thus write the $ds^2$ of Eq. (\ref
{ds2 3}) as
\begin{eqnarray}
\fl
ds^2=\sum_a\left\{ \left[ (\delta x_a)^2+(\delta y_a)^2-2\tau _a(\delta
z_a)(\delta x_a-\delta y_a-\delta z_a)\right] +\frac{\left( d\tau _a\right)
^2}{2\tau _a\rho _a}\right\} 
\nonumber \\
+2\sum_{a<b}\left\{\sum_{i=1}^4\left( \left( \delta a_i\right) _{ab}\right)
^2+\sum_{i=1}^4\left( \left( \delta s_i\right) _{ab}\right) ^2
\right.
\label{ds2 6}
\\[0.05in]
+2\sqrt{\rho _a\rho _b}\left[\left( \delta a_1\right) _{ab}\left( \delta
a_3\right) _{ab}+\left( \delta a_2\right) _{ab}\left( \delta a_4\right) _{ab}
+\left( \delta s_1\right) _{ab}\left( \delta s_3\right) _{ab}+\left( \delta
s_2\right) _{ab}\left( \delta s_4\right) _{ab}\right]
\nonumber \\[0.05in]
+2\sqrt{\tau _a\tau _b}\left[\left( \delta a_2\right) _{ab}\left( \delta
a_3\right) _{ab}+\left( \delta a_1\right) _{ab}\left( \delta a_4\right) _{ab}
+\left( \delta s_2\right) _{ab}\left( \delta s_3\right) _{ab}+\left( \delta
s_1\right) _{ab}\left( \delta s_4\right) _{ab}\right]\}.  
\end{eqnarray}

The metric tensor appearing in Eq. (\ref{arc element}) has a simple block
structure, consisting of $1\times 1$, $3\times 3$ and $4\times 4$ blocks
along the diagonal, as follows. There are $N$ $3$-dimensional blocks with
rows and columns labeled $\delta x_a$, $\delta y_a$, $\delta z_a$:
\begin{equation}
\begin{array}{cc}
 & \begin{array}{ccc}
   \delta x_a & \delta y_a & \delta z_a
   \end{array}
\\ $ $ \\
\begin{array}{c}
\delta x_a \\ 
\delta y_a \\ 
\delta z_a
\end{array}
&
\left[ 
\begin{array}{ccc}
1 & 0 & -\tau _a \\ 
0 & 1 & \tau _a \\ 
-\tau _a & \tau _a & 2\tau _a
\end{array}
\right]   
\end{array}
\label{block delta x,y,z a}
\end{equation}
and $N$ 1-dimensional blocks $1/2\rho _a\tau _a$ labeled by $d\tau _a$,
giving, altogether, the contribution
\begin{equation}
\prod _{a=1}^N\frac{2\tau _a(1-\tau _a)}{2\tau _a(1-\tau _a)}=1
\label{contr A to detg}
\end{equation}
to $\det g$. There are $N(N-1)/2$ $4$-dimensional blocks with rows and
columns labeled $\left( \delta s_1\right) _{ab}$, $\left( \delta s_2\right)
_{ab}$, $\left( \delta s_3\right) _{ab}$, $\left( \delta s_4\right) _{ab}$
(for $a<b$)
\[
\quad \quad \quad \quad 
\]
\begin{equation}
\begin{array}{cc}
 & \begin{array}{cccc}
\left( \delta s_1\right) _{ab} & \left( \delta s_2\right) _{ab} & \left(
\delta s_3\right) _{ab} & \left( \delta s_4\right) _{ab}
\end{array}
\\ $ $ \\
\begin{array}{c}
\left( \delta s_1\right) _{ab} \\ 
\left( \delta s_2\right) _{ab} \\ 
\left( \delta s_3\right) _{ab} \\ 
\left( \delta s_4\right) _{ab}
\end{array}
&
\left[ 
\begin{array}{cccc}
1 & 0 & \sqrt{\rho _a\rho _b} & \sqrt{\tau _a\tau _b} \\ 
0 & 1 & \sqrt{\tau _a\tau _b} & \sqrt{\rho _a\rho _b} \\ 
\sqrt{\rho _a\rho _b} & \sqrt{\tau _a\tau _b} & 1 & 0 \\ 
\sqrt{\tau _a\tau _b} & \sqrt{\rho _a\rho _b} & 0 & 1
\end{array}
\right]   
\end{array}
\label{block delta s1,2,3,4 ab}
\end{equation}
and also $N(N-1)/2$ $4$-dimensional blocks with rows and columns labeled $%
\left( \delta a_1\right) _{ab}$, $\left( \delta a_2\right) _{ab}$, $\left(
\delta a_3\right) _{ab}$, $\left( \delta a_4\right) _{ab}$ and identical to
the matrices of (\ref{block delta s1,2,3,4 ab}) giving, altogether, a
contribution
\begin{equation}
\left( \tau _a-\tau _b\right) ^4.  \label{contr B to detg}
\end{equation}

Multiplying (\ref{contr A to detg}) and (\ref{contr B to detg}) and taking
the square root, as required by Eq. (\ref{vol element}), we find the result
given in Eq. (\ref{w(tau)}).

\section{Appendix: The distribution of the conductance in the absence of direct
processes}
\label{w(T) no direct}

We derive here the two-channel conductance distribution of Eqs. (\ref{w(g)
N=2 beta=1}) and (\ref{w(g) N=2 beta=2}) and the behavior of that
distribution for arbitrary $N$ in the range $0<T<1$, as given by 
Eq. (\ref{w(g) N arb}).

\begin{figure}[tb]
   \begin{center}
   \leavevmode
   \epsfxsize= 6.5cm
   \epsffile[70 250 550 750]{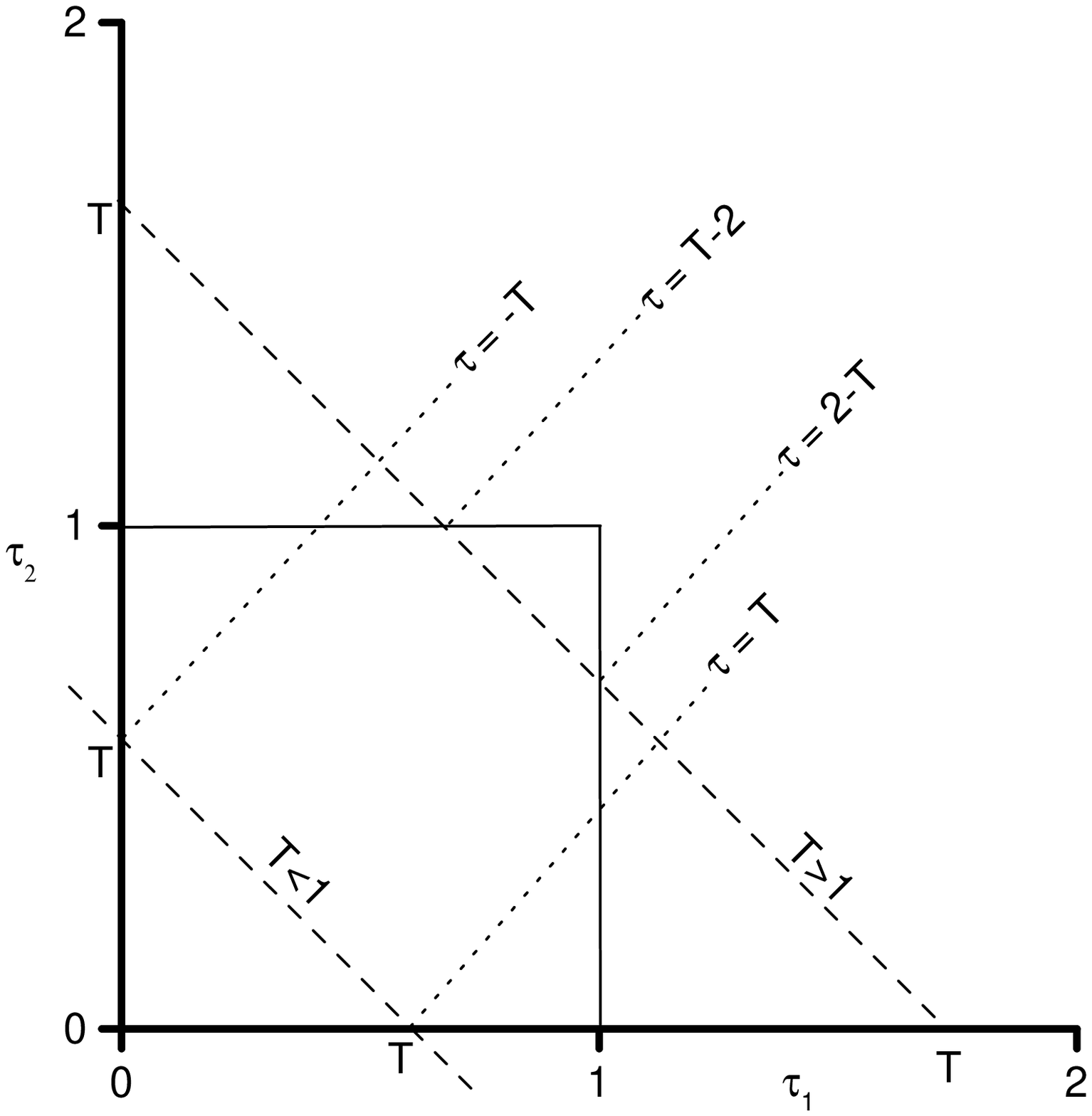}
   \end{center}
   \vspace*{-0.5cm}\caption{
   Region of integration for finding the distribution $w(T)$ 
   in the $N=2$ case.
   }
   \label{appB_regionint} 
\end{figure}

For the two-channel distribution we have to perform the integral 
(\ref{def of w(T)}) in the
two-dimensional space $\tau _1$, $\tau _2$, inside the square  $0<\tau _1<1$, 
$0<$ $\tau _2<1$, indicated in Fig. \ref{appB_regionint}. 
The change of variables
\begin{equation}
T=\tau _1+\tau _2 , \qquad 
\tau =\tau _1-\tau _2  
\label{Ttau12}
\end{equation}
will be found advantageous. As shown in Fig. \ref{appB_regionint}, 
when $0<T<1$ the variable $\tau $ varies in the interval $-T<\tau <T$, 
whereas when $1<T<2$ we have $T-2<\tau <2-T$. 

\subsection{Derivation of Eq.(\protect\ref{w(g) N=2 beta=2}).}

From Eq. (\ref{w(tau)}) for the invariant measure we find, for two
channels
\begin{equation}
P^{(2)}(\tau _1,\tau _2)=C\left( \tau _1-\tau _2\right) ^2.
\label{w(tau) beta=2 N=2}
\end{equation}
The normalization constant is found from
\begin{equation}
\fl
1=C\int \int_0^1\left( \tau _1-\tau _2\right) ^2d\tau _1d\tau _2
=\frac C2\left[ \int_0^1dT\int_{-T}^T
d\tau \cdot \tau ^2+\int_1^2dT
\int_{T-2}^{2-T}d\tau \cdot \tau^2 \right] =\frac C6,
\end{equation}
so that $C \smeq 6$. 

The distribution of the spinless conductance is thus
\begin{equation}
w^{(2)}(T)=3\left[\int_0^1dT^{\prime }\delta (T-T^{\prime })
\int_{-T^{^{\prime
}}}^{T^{\prime }}d\tau \cdot \tau ^2
+\int_1^2dT^{\prime }\delta (T-T^{\prime })\int_{T^{^{\prime
}}-2}^{2-T^{\prime }}d\tau \cdot \tau ^2 \right] ,
\label{w(T) N=2 beta=2 1}
\end{equation}
so that
\begin{equation}
w^{(2)}(T<1)=2T^3 , \qquad
w^{(2)}(1<T<2)=2(2-T)^3.  \label{w(T) N=2 beta=2 1<T<2}
\end{equation}
These two equations can be condensed in the result of Eq. (\ref{w(g) N=2
beta=2}).

\subsection{Derivation of Eq. (\protect\ref{w(g) N=2 beta=1})}

From Eq. (\ref{w(tau)}) for the invariant measure we find, for two
channels
\begin{equation}
P^{(1)}(\tau _1,\tau _2)=C\frac{\left| \tau _1-\tau _2\right| }{\sqrt{\tau
_1\tau _2}}.  \label{w(tau) beta=1 N=2}
\end{equation}
The normalization constant is found from
\begin{equation}
\fl
1=C\int \int_0^1\frac{\left| \tau _1-\tau _2\right| }{\sqrt{\tau _1\tau _2}}%
d\tau _1d\tau _2
\end{equation}
\[
=\frac C2\left[ \int_0^1dT\int_{-T}^{T}
d\tau\frac{2\left| \tau \right| }{\sqrt{T^2-\tau^2}}+
\int_1^2dT\int_{T
-2}^{2-T}d\tau \frac{2\left| \tau \right| }{%
\sqrt{T^2-\tau^2}}\right] 
=\frac 43C,
\]
so that $C \smeq 3/4$. 

The distribution of the spinless conductance is thus
\begin{eqnarray}
\fl
w^{(1)}(T)=\frac 38[\int_0^1dT^{\prime }\delta (T-T^{\prime
})\int_{-T^{^{\prime }}}^{T^{\prime }}d\tau \frac{2\left| \tau
\right| }{\sqrt{(T^{\prime })^2-\tau^2}}
\nonumber \\[0.1in]
+\int_1^2dT^{\prime }\delta (T-T^{\prime })\int_{T^{^{\prime
}}-2}^{2-T^{\prime }}d\tau \frac{2\left| \tau \right| }{%
\sqrt{(T^{\prime })^2-\tau ^2}},  \label{w(T) N=2 beta=1 1}
\end{eqnarray}
so that
\begin{equation}
w^{(1)}(T<1)=\frac 32T  , \qquad
w^{(1)}(1<T<2)=\frac 32\left[ T-2\sqrt{T-1}\right] ,
\label{w(T) N=2 beta=1 1<T<2}
\end{equation}
which are Eqs.  (\ref{w(g) N=2 beta=1}) we wanted to prove.

\subsection{Derivation of Eq. (\protect\ref{w(g) N arb})}

The integration region selected by the $\delta $ function in 
Eq. (\ref{def of w(T)}) is $N-1$
dimensional. In the particular case $T<1$, each of the $\tau _a$'s 
($a \smeq 1,\ldots,N$) varies, over that region, in the interval $0<\tau
_a<T$. This is clearly illustrated in Fig. \ref{appB_regionint}
for $N \smeq 2$, where the relevant
region is the segment extending from $(\tau _1,\tau _2) \smeq (0,T)$ to 
$(\tau _1 ,\tau _2) \smeq (T,0)$. For $T<1$ the corners of the $N$-dimensional
hypercube are not relevant; they become relevant for $T>1$. We can thus write, 
for $T<1$
\begin{equation}
w_N^{(\beta )}(T<1)
=C_N^{(\beta )}\int_0^T\cdot \cdot \cdot \int_0^T\delta \left( T-
\sum_{a=1}^N\tau _a\right) \prod _{a<b}\left| \tau _a-\tau _b\right|
^\beta \prod _c\tau _c^{\frac{\beta -2}2}\prod _id\tau _i
\label{w(T), N, beta, T<1}
\end{equation}
Introducing the new variables $\sigma _a \smeq \tau _a/T$
($a \smeq 1,\cdot \cdot \cdot N$) which vary in the interval $(0,1)$, we have
\begin{eqnarray}
\fl
\hspace*{-0.45cm}
w_N^{(\beta )}(T<1)
=C_N^{(\beta )}\int_0^1 \! \cdots \int_0^1\delta \left( T-
T\sum_{a=1}^N\sigma _a\right) T^{\frac{N(N-1)}2\beta }\prod
_{a<b}\left| \sigma _a-\sigma _b\right| ^\beta T^{N\frac{\beta -2}2}\prod
_c\sigma _c^{\frac{\beta -2}2}T^N\prod _id\sigma _i
\nonumber \\[0.1in]
=C_N^{(\beta )}T^{\beta \frac{N^2}2-1}\int_0^1 \! \cdots
\int_0^1\delta \left( 1-\sum_{a=1}^N\sigma _a\right) \prod
_{a<b}\left| \sigma _a-\sigma _b\right| ^\beta \prod _c\sigma _c^{\frac{%
\beta -2}2}\prod _id\sigma _i,  \label{w(T), N, beta, T<1, 1}
\end{eqnarray}
which behaves as the power of $T$ indicated in Eq. (\ref{w(g) N arb}).


\begin{thebibliography}{99}
\bibitem{mucciolo-private}
The same effect has been noticed in the context of random matrix 
Hamiltonians: one can have Gaussian ensemble eigenvalue statistics while
not having Porter-Thomas wavefunction statistics (E. Mucciolo, private
communication).
\bibitem{superpose intervals}
While $\langle S \rangle$ varies slightly over the full energy range used,
the distribution of $T$ implied by $\langle S \rangle$ does not vary. Thus,
the theoretical curve shown in Fig. \protect\ref{num_presence} is the 
average of the prediction of Poisson's kernel for each of the four 
intervals of $50$ energies. The numerical result is the histogram 
of all of the data.
\bibitem{correct exp var}
To compensate for the thermal smearing in the experiment, we estimate
the energy correlation length $E_c$ from the data given and then multiply
the experimental variance by $3.5 kT/E_c$. For \protect\cite{Marcus95},
this produces a factor of $\sim 2.6$-$4.5$.
\bibitem{notPoisKern}
For the simple structures of section \protect\ref{simple structures}, in
addition to comparing the numerical results to the invariant-measure theory,
we also compared to Poisson's kernel. The agreement was also poor in this
case because the statistics were not stationary so that an optical $S$-matrix
could not be unambiguously defined. In contrast, the structures of section
\protect\ref{numerics with d.p.} have smaller openings: this causes two
widely separated time scales, and so Poisson's kernel is a good description.
\end{thebibliography}
\end{document}